\documentclass[aps, twocolumn, showpacs, letterpaper]{revtex4}

\usepackage{amsmath}
\usepackage{amssymb}
\usepackage{xspace}
\usepackage{url}

\usepackage{ifpdf}
\ifpdf
\fi

\newcommand{\eg}{{e.g.,\/}\xspace}
\newcommand{\ie}{{i.e.,\/}\xspace}

\newcommand{\gap}{\mbox{}}

\newcommand{\eq}[1]{(\ref{#1})}
\newcommand{\Eq}[1]{Eq.~(\ref{#1})}
\newcommand{\Eqs}[1]{Eqs.~(\ref{#1})} 
 
\newcommand{\Ref}[1]{Ref.~\cite{#1}}
\newcommand{\Refs}[1]{Refs.~\cite{#1}}
\newcommand{\Sec}[1]{Sec.~\ref{#1}}
\newcommand{\Secs}[1]{Secs.~\ref{#1}}

\newcommand{\msection}[1]{\textit{#1.}\ ---\ }

\newcommand{\mc}[1]{\mathcal{#1}}
\newcommand{\mcc}[1]{\mathfrak{#1}}
\newcommand{\msf}[1]{\mathsf{#1}}

\renewcommand{\vec}[1]{{\boldsymbol{\rm #1}}}
\newcommand{\favr}[1]{\langle #1 \rangle}
\newcommand{\oper}[1]{\hat{\vec{#1}}}
\newcommand{\pd}{\partial}
\newcommand{\del}{\nabla}
\newcommand{\kpt}[1]{{\kern #1 pt}}
\newcommand{\const}{\text{const}}

\begin{document}
\title{Axiomatic geometrical optics, Abraham-Minkowski controversy,\\and photon properties derived classically}
\author{I.~Y. Dodin and N.~J. Fisch}
\affiliation{Princeton Plasma Physics Laboratory, Princeton, New Jersey 08543, USA}
\date{\today}

% 03.50.De -- Classical electromagnetism, Maxwell equations
% 42.50.Wk -- Mechanical effects of light on material media, microstructures and particles
% 42.15.-i -- Geometrical optics
% 03.50.-z -- Classical field theories

% Other:
% 45.20.-d -- Formalisms in classical mechanics
% 45.20.Jj -- Lagrangian mechanics
% 14.70.Bh -- Photons, properties of
% 45.30.+s -- General linear dynamical systems
% 42.25.Bs -- Wave propagation, transmission and absorption
% 42.50.-p -- Quantum optics
% 42.50.Tx -- Optical angular momentum and its quantum aspects
% 37.10.Vz -- Mechanical effects of light on atoms, molecules, and ions 


\begin{abstract}
By restating geometrical optics within the field-theoretical approach, the classical concept of a photon (and, more generally, any elementary excitation) in arbitrary dispersive medium is introduced, and photon properties are calculated unambiguously. In particular, the canonical and kinetic momenta carried by a photon, as well as the two corresponding energy-momentum tensors of a wave, are derived from first principles of Lagrangian mechanics. As an example application of this formalism, the Abraham-Minkowski controversy pertaining to the definitions of these quantities is resolved for linear waves of arbitrary nature, and corrections to the traditional formulas for the photon kinetic energy-momentum are found. Several other applications of axiomatic geometrical optics to electromagnetic waves are also presented.
\end{abstract}

\maketitle

\section{Introduction}
\label{sec:intro}

%--------------------------------------
\subsection{Motivation}

The discussion about how to define the momentum and the angular momentum of a photon in dispersive medium (PDM), and even simply of a classical wave, recurs in literature periodically during the last hundred years. The recent burst of theoretical \cite{ref:barnett10, ref:philbin12, ref:rikken12, ref:crenshaw12, arX:wang12, ref:ramos11, ref:greenshaw11, ref:he11, ref:liu11, ref:brevik11, ref:philbin11, ref:jimenez11, ref:yu11, arX:nakamura11, arX:ravndal11, arX:ojima11, arX:wang11, arX:zhang11, ref:mansuripur10, ref:shevchenko10, ref:saldanha10, ref:bradshaw10, ref:veselago10, ref:pfeifer09, ref:hinds09, ref:mansuripur09, arX:wang09, ref:padgett08, ref:scullion08, ref:kemp07, ref:dereli07a, ref:dereli07b, ref:mansuripur07, ref:leonhardt06, ref:leonhardt06b, ref:stalinga06a, ref:stalinga06b, ref:loudon05, ref:mansuripur05, ref:milonni05, ref:garrison04, ref:feigel04, ref:lopez04, ref:mansuripur04, ref:tiggelen04, ref:padgett03, ref:obukhov03, ref:loudon02, ref:loudon97, ref:labardi96, ref:kristensen94, ref:baxter93, ref:nelson91} and experimental \cite{ref:wang11, ref:she08, ref:campbell05} publications indicates both an abiding interest in the problem and, apparently, a lack of consensus or certainty about what the correct answer is. The traditional arguments can be found in reviews like \Refs{ref:barnett10b, ref:kemp11, ref:milonni10, ref:baxter10, ref:pfeifer07, ref:loudon04, ref:brevik79} and references therein, too numerous to be listed in this paper. Let us mention only briefly that two alternative forms of the PDM momentum are adopted most commonly:
\begin{gather}\label{eq:ma}
p_{\rm M} = \hbar\omega n_{\rm p}/c, \quad p_{\rm A} = \hbar\omega/(n_{\rm g} c), 
\end{gather}
which are known, respectively, as the Minkowski interpretation and the Abraham interpretation \cite{foot:history}. (Here $\omega$ is the frequency, $c$ is the speed of light, and $n_{\rm p} = c/v_{\rm p}$ and $n_{\rm g} = c/v_{\rm g}$ are the refraction indexes associated with, correspondingly, the phase velocity $v_{\rm p}$ and the group velocity $v_{\rm g}$; for the two associated angular momenta see \Ref{ref:padgett03}.) Since both have supporting theoretical and experimental evidence \cite{ref:barnett10}, the question about which of the two interpretations is ``more correct'' has been controversial.

A resolution to this Abraham-Minkowski controversy (AMC) was proposed recently in \Ref{ref:barnett10}. It was argued there that \textit{both} interpretations are correct; namely, $p_{\rm M}$ can be attributed as the canonical momentum, and $p_{\rm A}$ can be attributed as the kinetic momentum of a photon. Yet, strictly speaking, the argument of \Ref{ref:barnett10} applies only to the case of a nonrelativistic solid dielectric. The subsequent generalization in \Ref{ref:ramos11} is not quite complete either; for example, the latter neglects electrostriction and magnetostriction, kinetic effects, and spatial dispersion, and also attributes $v_{\rm g}$ entirely to the Poynting flux, in disagreement with a textbook theorem [\Eq{eq:vgsk}]. Thus, a quantitative relativistic theory is still lacking that would correct the existing understanding of PDM, and \Eqs{eq:ma} in particular. The purpose of this paper is to resolve these issues in a consistent manner and, through that, formulate a comprehensive asymptotic theory of linear waves of arbitrary nature.

%--------------------------------------
\subsection{Field-theoretical approach}

Before photon properties can be calculated, the PDM itself must be defined unambiguously. (In particular this means that, contrary to the common presumption, the PDM properties cannot simply be inferred from experiment without a theory of what is being inferred.) Second of all, the definition should not be expected to originate from electromagnetism, because the concept of a photon, and even of $v_{\rm p}$ and $v_{\rm g}$ that enter \Eqs{eq:ma}, is not embedded in Maxwell's equations \textit{per~se}. On the other hand, the photon concept is neither entirely of quantum nature \cite{ref:mcdonald85}, and mechanical properties of quantum radiation (dipole force, radiation pressure, cooling effects on atoms, etc) are consistently shown to have direct classical analogs \cite{arX:mycoin, my:nlinphi, my:kchi,my:manley, my:invited}. It then stands to reason that an abstract classical calculation could resolve the AMC, generalizing \Eqs{eq:ma}, without assuming a specific underlying physical system whatsoever.

To understand what the right framework is for such a calculation, notice that introducing a photon implies that the frequency $\omega$ and the wave vector $\vec{k}$ are well defined. These are exactly the validity conditions of the asymptotic theory commonly known as geometrical optics (GO). (The term ``optics'' here means only that the theory deals with sufficiently large $\omega$ and $k$; \ie waves need not be electromagnetic.) Although usually defined through rays and wave equations \cite{book:landau2, book:landau8, book:kravtsov, ref:weinberg62, ref:kravtsov74, ref:bernstein75, ref:bernstein77, ref:bornatici00, ref:bornatici03}, the most fundamental, axiomatic GO is an abstract field theory that applies to any field having a Lagrangian density of a specific form [\Eq{eq:L}; dissipative effects can also be added, \Sec{sec:dissw}]. Just like Newton's laws of particle motion hold, with obvious exceptions, independently of specific forces acting on particles, the basic GO equations are then invariant to the wave nature \cite{foot:such}, and the wave properties can be derived in general. Hence, axiomatic GO should resolve the AMC automatically and transparently.

%--------------------------------------
\subsection{Outline}
\label{sec:outline}

Here we aim to apply the GO formalism toward deriving PDM general properties deductively using nothing more than first principles of classical mechanics. In doing so, we draw on the Lagrangian field theory as elaborated in plasma physics and hydrodynamics during the last fifty years \cite{ref:sturrock58, ref:sturrock60b, ref:toupin60, ref:sturrock61, ref:whitham65, book:whitham, ref:dougherty70, ref:dewar77, ref:lighthill65b, ref:garrett67, ref:bretherton69, ref:hayes70, ref:dewar70, ref:jones71, ref:dewar72c, ref:jones73, ref:dewar73, ref:hayes73, ref:melrose81, ref:heintzmann83, ref:takahashi86, ref:kentwell87, ref:kulsrud92, ref:brizard93, ref:brizard95, my:bgk, my:acti, my:actii, my:actiii}. Since this literature, sadly, remains unknown within the mainstream approach to the AMC (with few exceptions), the general formalism of axiomatic GO will also be restated.

Specifically, below we do the following:
\begin{itemize}
\item[(a)] formulate from first principles a comprehensive theory of axiomatic GO, extending and expanding on existing results in applications to waves of arbitrary nature (not just electromagnetic waves);
\item[(b)] explain how the wave canonical energy-momentum tensor (EMT) is related to the photon properties in the Minkowski interpretation (here and further photons are understood as any elementary excitations, not just light quanta; \Secs{sec:covar} and \ref{sec:quantum});
\item[(c)] introduce the wave angular momentum and photon (plasmon, phonon, polariton, etc.) spin within axiomatic GO and calculate it explicitly for cylindrical beams, also of arbitrary nature;
\item[(d)] derive the effect of local linear dissipation; 
\item[(e)] unambiguously define the wave kinetic EMT and calculate it explicitly for isotropic relativistic fluids (with striction effects included); 
\item[(f)] calculate the associated energy, momentum, and angular momentum per photon (plasmon, phonon, polariton, etc.); show that the traditional, Abraham's formulas are reproduced as a limiting case;
\item[(g)] illustrate how the properties of electromagnetic waves, \textit{considered as a special case}, can be inferred deductively within axiomatic GO;
\item[(h)] finally, responding to the questions posed in \Ref{ref:leonhardt06b}, we clarify the applicability of Minkowski's and Abraham's formulas for electromagnetic waves in various media (including cold, warm, and relativistic) and present examples.
\end{itemize}
Note that, in parts (a) and (b), which correspond to \Secs{sec:general}-\ref{sec:noether}, we mostly repeat known arguments, published previously, \eg in \Refs{ref:sturrock61, ref:whitham65, book:whitham, ref:dougherty70, ref:dewar77}. Also keep in mind that, in application to specific media, the problem of finding both canonical and kinetic EMT of a classical electromagnetic wave was solved comprehensively in \Ref{ref:dewar77}, which, while known within the plasma physics community, seems to remain unknown to the general readership. The difference between \Ref{ref:dewar77} and our paper is that we use different machinery to arrive at results that are, in certain aspects, more general and, as a consequence, more concise and transparent. In particular, \textit{our paper is not about electromagnetism but rather about basic physics of waves}, so we need not specify the wave nature; also we derive photon properties, and allow for dissipation, complementing \Ref{ref:dewar77} on these issues.

The paper is organized as follows. In \Sec{sec:notation} we introduce the notation used throughout the text. In \Sec{sec:general} we describe general GO waves, including nonlinear waves, in arbitrarily curved spacetime and also in the Minkowski spacetime as a particular case. In \Sec{sec:linear} we reduce the theory further to describe linear waves and explain how the Minkowski representation is recovered; in particular, the wave angular momentum and dissipative effects are discussed. In \Sec{sec:amc} we introduce the wave kinetic EMT and reproduce the traditional formulas for the corresponding photon quantities as a limiting case. In \Sec{sec:em} we consider, as an example, how the specific properties of electromagnetic waves flow deductively from the general theory. \Sec{sec:disc} explores ramifications of our findings and summarizes our main results. Auxiliary calculations are presented in Appendix.

%%%%%%%%%%%%%%%%%%%%%%%%%%%%%%%%%%%%%%%
\section{Notation}
\label{sec:notation}

The following notation will be assumed below. We use the symbol $\doteq$ for definitions. Greek indexes span from 0 to 3 and refer to coordinates in spacetime, $x^\alpha$. In particular, for the Minkowski spacetime we adopt $x^0 \doteq ct$, where $t$ is time. Hence the Lorentz transformation matrix, ${\Lambda^\alpha}_\beta \doteq \pd x^\alpha/\pd x'^\beta$, is given by
\begin{gather}
{\Lambda^0}_0 = \gamma, \quad {\Lambda^0}_i = \gamma v_i/c, \quad {\Lambda^i}_0 = \gamma v^i/c, \notag\\
{\Lambda^i}_j = \delta^i_j + (\gamma - 1)v^iv_j/v^2, \label{eq:lorentz}
\end{gather}
where $v^i$ is the velocity of the ``primed'' reference frame with respect to the laboratory frame, and $\gamma \doteq (1 - v^2/c^2)^{-1/2}$. Latin indexes $i$, $j$, and $l$ span from 1 to 3 and refer to spatial coordinates,~$x^i$. Spatial vectors are denoted with bold, $\vec{X}$; spatial tensors are also marked with hat, $\oper{T}$; symbols like $\vec{X}\vec{Y} \doteq \oper{Z}$ stand for spatial dyadics, $Z^{ij} = X^i Y^j$; the symbol $\oper{1}$ denotes the unit spatial tensor; besides, the three-tensor
\begin{gather}
\oper{\Lambda} \doteq \oper{1} + \frac{\gamma - 1}{v^2}\,\vec{v}\vec{v}
\end{gather}
is the spatial part of ${\Lambda^\alpha}_\beta$. Summation over repeating indexes will also be implied; \eg $X^i Y_i \equiv \sum_{i=1}^3 X^i Y_i$. 

Latin indexes (excluding $i$, $j$, $l$, and non-bold roman, as in $n_{\rm p}$) denote partial derivatives with respect to the corresponding variables; \eg for $f \doteq f(a, \omega, \vec{k}; t, \vec{x})$, the symbol $f_{\vec{x}}$ denotes the derivative (gradient) with respect to the last argument, $\vec{x}$. In addition to those, we also introduce ``full'' temporal and spatial derivatives, $\pd_t \msf{X}\equiv \pd \msf{X}/\pd t$ and $\pd_i \msf{X}\equiv \pd \msf{X}/\pd x^i$, which treat \textit{all} arguments of (any) $\msf{X}$ as functions of, correspondingly, $t$ and $x^i$. For instance, for the above $f$, one has
\begin{gather}
\pd_t f = f_a\, \pd_t a + f_\omega\, \pd_t \omega + f_{k_i} \pd_t k_i + f_t, \\
\pd_i f = f_a\, \pd_i a + f_\omega\, \pd_i \omega + f_{k_j} \pd_i k_j + f_{x_i}, 
\end{gather}
while $\pd_t a(t, \vec{x}) = a_t(t, \vec{x})$, etc. The symbol $\del$ denotes the associated full covariant derivative; \eg $\del_i f = \pd_i f$ is the full gradient of the scalar $f$, and $\del \cdot \vec{F}$ is the full divergence of the vector~$\vec{F}$,
\begin{gather}
\del \cdot \vec{F} = \frac{1}{\sqrt{\eta}}\,\frac{\pd}{\pd x^i}\,\big(\sqrt{\eta}\,F^i\big), 
\end{gather}
where $\eta \doteq \text{det}\,\eta_{ij}$, and $\eta_{ij} = \eta_{ji}$ is the spatial metric. [In Cartesian coordinates, Euclidean space has $\eta_{ij} = \eta^{ij} = \text{diag}\,(1,1,1)$, so $\eta = 1$.] The symbol~$_{, \alpha}$ denotes the analogous (to $\pd_i$) full derivative with respect to $x^\alpha$, and~$_{; \alpha}$~denotes the analogous full covariant derivative. For example, the four-divergence~is
\begin{gather}
{F^{\alpha}}_{;\alpha} = \frac{1}{\sqrt{g}}\,\frac{\pd}{\pd x^\alpha}\,\big(\sqrt{g}\,F^\alpha\big), 
\end{gather}
where $g \doteq - \text{det}\, g_{\mu\nu}$, and $g_{\mu\nu} = g_{\nu\mu}$ is the spacetime metric. For introduction to the tensor notation and index manipulation rules in particular, see \Refs{my:metric, book:weinberg, book:landau2}.

Some specific symbols are also summarized in Table~\ref{tab:table}, and the abbreviations used in the text are as follows:\\[10pt]
\begin{tabular}{@{\quad} r@{\quad -- \quad} l @{\quad}}
ACT & action conservation theorem,\\
AMC & Abraham-Minkowski controversy,\\
EMT & energy-momentum tensor,\\
GO & geometrical optics,\\
PDM & photon in dispersive medium,\\
SAM & spin angular momentum,\\
WMS & ``wave + medium'' system.\\
\end{tabular}\\

\begin{table*}
\caption{Summarized here is some notation adopted for wave variables (``ponder.'' stands for ponderomotive; the integral quantities are obtained by integrating the corresponding densities over the spatial volume, $dV \equiv \sqrt{\eta}\,d^3x$). The rest of the notation is explained in \Sec{sec:notation} and throughout the text.}
\label{tab:table}
\begin{center}
{
\begin{tabular}{l@{\qquad} c@{\quad}c@{\quad}c @{\qquad}c@{\qquad} c@{\quad}c}
\hline\hline
&
\multicolumn{3}{c}{per unit spatial volume}\qquad & 
integral & 
\multicolumn{2}{c}{per photon}\\
\hline
& canonical & kinetic & ponder. & - & canonical & kinetic\\
\hline
number of photons & $\mc{N}$ & - & - & $N$ & $1$ & -\\
action & $\mc{I}$ & - & - & $I$ & $\hbar$ & - \\
energy & $\mc{E}$ & $\varepsilon$ & $\Delta\varepsilon$ & - & $H$ & $h$\\
momentum & $\vec{\mc{P}}$ & $\vec{\rho}$ & $\Delta\vec{\rho}$ & - & $\vec{P}$ & $\vec{p}$\\
angular momentum & $\vec{\mc{M}}$ & $\vec{\mu}$ & $\Delta\vec{\mu}$ & - & $\vec{M}$ & $\vec{m}$\\
photon flux & $\vec{\mc{G}}$ & - & - & - & - & - \\
action flux & $\vec{\mc{J}}$ & - & - & - & - & - \\
energy flux & $\vec{\mc{Q}}$ & $\vec{\vartheta}$ & - & - & - & - \\
momentum flux & $\oper{\Pi}$ & $\oper{\pi}$ & - & - & - & - \\
energy-momentum tensor & $\mc{T}$ & $\tau$ & $\Delta\tau$ & - & $T$ & - \\
wave Lagrangian & $\mcc{L}$ & - & - & $\msf{L}$ & - & - \\
\hline\hline
\end{tabular}
}
\end{center}
\end{table*}

%%%%%%%%%%%%%%%%%%%%%%%%%%%%%%%%%%%%%%%
\section{General waves}
\label{sec:general}

%--------------------------------------
\subsection{Covariant formulation}
\label{sec:covar}

First, let us consider a general nondissipative wave described by some action integral $S = \int \mcc{L} \,\sqrt{g}\,d^4 x$, where
\begin{gather}
\sqrt{g}\,d^4 x \equiv \sqrt{g}\,dx^1 dx^2 dx^3 dx^4
\end{gather}
is an invariant volume element in spacetime, and the four-scalar $\mcc{L}$ is the Lagrangian density. Since the action of the underlying medium is not included here, no invariance requirements on $\mcc{L}$ are imposed. Instead, we assume that the wave structure remains fixed (albeit not necessarily sinusoidal), so the wave is fully described by some canonical phase $\theta$, which will be understood as a scalar field $\theta(x^\nu)$, and $a = a(x^\nu)$, which is an arbitrary measure of the wave local amplitude \cite{foot:veca}. We also assume that the envelope evolves on spacetime scales large compared to those of local oscillations. On such time scales, it is only the average Lagrangian density that contributes to $S$, so one can adopt that $\mcc{L}$ does not depend on $\theta$ explicitly. Instead, $\mcc{L}$ must depend on the phase four-gradient,
\begin{gather}\label{eq:kalpha}
k_\mu \doteq \theta_{, \mu},
\end{gather}
which is the generalized ``wave vector'' (actually, a four-covector here), obviously having zero four-curl,
\begin{gather}\label{eq:consist}
k_{\mu;\nu} - k_{\nu;\mu} = k_{\mu,\nu} - k_{\nu,\mu} = \theta_{,\mu\nu} - \theta_{,\nu\mu} = 0.
\end{gather}
[Equation \eq{eq:consist} is known as the consistency relation.] Besides that, $\mcc{L}$ must depend on $a$; yet the dependence on the amplitude gradients $a_{,\nu}$ is negligible in the GO limit. Thus, allowing also for slow parametric dependence on the spacetime coordinates $x^\nu$, we postulate
\begin{gather}\label{eq:L}
\mcc{L} = \mcc{L}(a, k_\mu; x^\nu),
\end{gather}
which as well can be considered as the \textit{definition} of the GO approximation. Hence wave equations are inferred using the least action principle, namely, as follows. 

First, let us consider variation of $S$ with respect to the wave amplitude $a$. Since $\delta_a S = \int \mcc{L}_a \,\delta a\,\sqrt{g}\,d^4 x$ for any $\delta a$, the requirement $\delta_a S = 0$ leads to
\begin{gather}\label{eq:ndr}
\mcc{L}_a = 0.
\end{gather}
Equation \eq{eq:ndr} can be understood as the wave dispersion relation, and it is generally nonlinear, \ie may retain essential dependence on~$a$ (see, \eg \Refs{my:bgk, my:actii}).

Second, let us consider variation of $S$ with respect to the wave phase $\theta$ \cite{foot:xik}. Due to \Eq{eq:kalpha} and the fact that $\mcc{L}$ does not depend on $\theta$ explicitly, for any $\delta \theta$ one has
\begin{multline}
%This \textstyle MUST be kept as is!
\delta_\theta S = \textstyle \int \mcc{L}_{k_\mu}\,\delta \theta_{,\mu}\,\sqrt{g}\,d^4 x \\
= \textstyle \int \big[(\sqrt{g}\,\mcc{L}_{k_\mu}\,\delta \theta)_{,\mu} - (\sqrt{g}\,\mcc{L}_{k_\mu})_{,\mu}\, \delta \theta \big]d^4 x\\
= - \textstyle \int (\mcc{L}_{k_\mu})_{;\mu}\, \delta \theta\,\sqrt{g}\,d^4 x,\notag
\end{multline}
where we used the fact that the wave field vanishes at infinity, so $\int (\ldots)_{,\mu}\,d^4x = 0$. Thus, the requirement $\delta_\theta S = 0$ yields that the four-divergence of the action flux density $\mc{J}^\mu \doteq - \mcc{L}_{k_\mu}$ is zero \cite{foot:manley},
\begin{gather}\label{eq:actgen}
{\mc{J}^\mu}_{;\mu} = 0,
\end{gather}
which is called the action conservation theorem (ACT). Since the ACT has the form of a continuity equation, one can treat $\mc{G}^\mu \doteq \mc{J}^\mu/\hbar$ as the flux density of some fictitious quasiparticles, or ``photons''. (In application to specific waves, one can as well think of plasmons, phonons, polaritons, or any other elementary excitations instead. See also \Sec{sec:quantum}.) However, remember that, within our classical description, it is only the product $\hbar \mc{G}^\mu$ that has an explicit physical meaning, so the actual value of $\hbar$ will be irrelevant for our purposes.

Finally, let us also introduce the wave EMT as follows. Consider the (generally asymmetric) tensor
\begin{gather}\label{eq:emt2}
{\mc{T}_\alpha}^\beta \doteq k_{\alpha}\mc{J}^\beta + \delta^\beta_\alpha \mcc{L}.
\end{gather}
The divergence of ${\mc{T}_\alpha}^\beta$ equals
\begin{align}
{{\mc{T}_{\alpha}}^\beta}_{; \beta} 
& = k_{\alpha;\beta}\mc{J}^\beta + k_\alpha {\mc{J}^\beta}_{;\beta} + \delta^\beta_\alpha (\mcc{L}_a\,a_{;\beta} +
\mcc{L}_{k_\lambda} k_{\lambda;\beta} + \mcc{L}_{x^\beta}) \notag\\
& = k_{\alpha;\beta}\mc{J}^\beta + \delta^\beta_\alpha (\mcc{L}_{k_\lambda} k_{\lambda;\beta} + \mcc{L}_{x^\beta}) \notag\\
& = k_{\alpha;\beta}\mc{J}^\beta - k_{\lambda;\alpha}\mc{J}^\lambda + \mcc{L}_{x^\alpha} \notag\\
& = k_{\alpha;\beta}\mc{J}^\beta - k_{\alpha; \lambda}\mc{J}^\lambda + \mcc{L}_{x^\alpha} \notag\\
& = \mcc{L}_{x^\alpha},\label{eq:emteq}
\end{align}
where we used \Eqs{eq:L}-\eq{eq:actgen}. This tensor is then associated with the conservation law, ${{\mc{T}_{\alpha}}^\beta}_{; \beta} = 0$, yielded when the system is translationally invariant in spacetime (\ie when the four-force is zero, $\mcc{L}_{x^\alpha} = 0$). Hence, ${\mc{T}_{\alpha}}^\beta$ is a true canonical EMT \cite{foot:sign}, as one could also infer from the standard definition that is based on Noether's theorem \cite{foot:noether}. However, notice that, in contrast with the fundamental theorem of the vacuum field theory \cite[Sec.~32]{book:landau2}, $\mc{T}^{\alpha\beta}$ does not permit the usual \cite{ref:belinfante39, ref:belinfante40, ref:pauli41, ref:mclennan66, ref:dewar77, ref:takahashi86} symmetrization, since $\mcc{L}$ is not restricted by any invariance requirements \cite{ref:sturrock58}.  (Yet see \Ref{book:hehl} for symmetrization via adopting an effective, ``optical'' metric.) In particular the very fact that a scalar field such as $\theta(x^\nu)$ yields an asymmetric EMT already proves the lack of Lorentz invariance \cite[Sec.~5.6]{book:logan}.

%--------------------------------------
\subsection{Application to the Minkowski spacetime}
\label{sec:mink}

From now on, we will assume the Minkowski spacetime with metric signature $(-,+,+,+)$; hence,
\begin{gather}\label{eq:metricgeta}
 g_{00} = g^{00} = - 1, \quad \eta_{ij} \doteq g_{ij}, \quad \eta = g.
\end{gather}
(Although the space is Euclidean, we will allow for curvilinear coordinates; thus, albeit flat, the spatial metric $\eta_{ij}$ can otherwise be arbitrary.) In this case, $k_\alpha = (- \omega/c, \vec{k})$, and $k^\alpha = (\omega/c, \vec{k})$, where
\begin{gather}\label{eq:omk}
\omega \doteq - \pd_t \theta, \quad \vec{k} \doteq \del \theta.
\end{gather}
Then \Eq{eq:consist} turns into the following set of equations:
\begin{gather}\label{eq:consistvec}
\pd_t \vec{k} + \del \omega = 0, \quad \del \times \vec{k} = 0.
\end{gather}
One may notice also that the latter equation here can be considered as the \textit{initial condition} for the former one, taking curl of which readily yields $\pd_t (\del \times \vec{k}) = 0$.

Accordingly, \Eq{eq:L} becomes
\begin{gather}\label{eq:LL}
\mcc{L} = \mcc{L}(a, \omega, \vec{k}; t, \vec{x}).
\end{gather}
The dispersion relation hence holds in the form \eq{eq:ndr}. The ACT can be rederived from \Eq{eq:LL} or it can be deduced from \Eq{eq:actgen} by substituting $\mc{J}^\alpha = (c\mc{I}, \vec{\mc{J}})$; either way, one gets (cf. \Refs{ref:whitham65, book:whitham})
\begin{gather}\label{eq:act2}
\pd_t \mc{I} + \del \cdot \vec{\mc{J}} = 0,
\end{gather}
where $\mc{I}$ is the action density, and $\vec{\mc{J}}$ is the action spatial flux density, introduced as follows:
\begin{gather}\label{eq:ii}
\mc{I} \doteq \mcc{L}_\omega, \quad \vec{\mc{J}} \doteq - \mcc{L}_\vec{k}.
\end{gather}
In particular, integration of \Eq{eq:act2} over the volume $dV \equiv \sqrt{\eta}\,d^3x$ yields conservation of the integral action,
\begin{gather}\label{eq:acttot}
I \doteq \int \mc{I}\, dV = \const.
\end{gather}
Introducing the photon density $\mc{N} \doteq \mc{I}/\hbar$ and the photon spatial flux density $\vec{\mc{G}} \doteq \vec{\mc{J}}/\hbar$, one can further rewrite \Eq{eq:act2} as $\pd_t \mc{N} + \del \cdot \vec{\mc{G}} = 0$, and \Eq{eq:acttot} will yield the photon conservation, $N \doteq \int \mc{N}\,dV = \const$. Also notice that both $I$ and $N$ are Lorentz invariants, as well-known to flow from the general (unlike, \eg in \Ref{ref:avron99}) properties of the continuity equation \cite[Sec.~2.6]{book:weinberg}. 

The elements of the (contravariant) EMT are now
\begin{gather}\notag
\mc{T}^{00} = \omega \mc{I} - \mcc{L}, \quad \mc{T}^{0i} = \omega \mc{J}^i/c, \\
\mc{T}^{i0} = ck^i \mc{I}, \quad \mc{T}^{ij} = k^i \mc{J}^j + \eta^{ij}\,\mcc{L}.
\end{gather}
In particular, \Eq{eq:emteq} yields
\begin{gather}\label{eq:e2}
\frac{\pd \mc{T}^{00}}{\pd t} + \frac{1}{\sqrt{\eta}}\,\frac{\pd}{\pd x^i}\, (c \mc{T}^{0i}\sqrt{\eta}) = w,
\end{gather}
which is a continuity equation for $\mc{T}^{00}$ with the right-hand side being $w \doteq g^{00}c \mcc{L}_{x^0} = - \mcc{L}_t$. Since the latter has the meaning of the canonical power source, $\mc{E} \doteq \mc{T}^{00}$ must be the wave canonical energy density, and $\mc{Q}^i \doteq c\mc{T}^{0i}$ must be the canonical energy flux density. Similarly,
\begin{gather}\label{eq:p2}
\frac{1}{c}\,\frac{\pd \mc{T}^{i0}}{\pd t} + \frac{1}{\sqrt{\eta}}\,\frac{\pd}{\pd x^j}\, (\mc{T}^{ij}\sqrt{\eta}) = f^i,
\end{gather}
which is a continuity equation for the three-vector $\mc{T}^{i0}/c$ with the right-hand side being $\vec{f} \doteq \mcc{L}_\vec{x}$. Since the latter has the meaning of the canonical momentum source, $\mc{P}^i \doteq \mc{T}^{i0}/c$ must be the wave canonical momentum density, and the (generally asymmetric) three-tensor $\Pi^{ij} \doteq \mc{T}^{ij}$ must be the canonical momentum flux density \cite{foot:mcin}. 

In summary, one then has
\begin{gather}\label{eq:block}
\mc{T}^{\alpha\beta} = \left(
\begin{array}{c @{\quad} c}
\mc{E} & \vec{\mc{Q}}/c\\[3pt]
c\vec{\mc{P}} & \oper{\Pi}
\end{array}
\right),
\end{gather}
where the individual blocks are given by
\begin{gather}
\mc{E} = \omega \mc{I} - \mcc{L}, \quad \vec{\mc{Q}} = \omega \vec{\mc{J}}, 
\notag\\
\vec{\mc{P}} = \vec{k}\mc{I}, \quad \oper{\Pi} = \vec{k} \vec{\mc{J}} + \mcc{L}\,\oper{1},\label{eq:eqppi}
\end{gather}
and \Eqs{eq:e2} and \eq{eq:p2} can be written as follows:
\begin{gather}
\pd_t \mc{E} + \del \cdot \vec{\mc{Q}} = w, \quad
\pd_t \vec{\mc{P}} + \del \cdot \oper{\Pi} = \vec{f}.\label{eq:EP}
\end{gather}
It is hence seen that the wave energy propagates at velocity $\vec{\mc{Q}}/\mc{E}$ that is generally different from the action flow velocity $\vec{\mc{J}}/\mc{I}$ [cf. \Eq{eq:act2}], and similarly for the momentum flow velocity. Moreover, those three turn out to be different from the velocities of \textit{information}, or the nonlinear group velocities, of which there can also be more than one. For an expanded discussion on this see \Refs{my:actiii, book:whitham} and references therein.

%%%%%%%%%%%%%%%%%%%%%%%%%%%%%%%%%%%%%%%
\section{Linear waves: Minkowski representation}
\label{sec:linear}

%--------------------------------------
\subsection{Basic equations}
\label{sec:linbasic}

Now let us consider a linear wave, \ie such that has $\omega(\vec{k}; t, \vec{x})$ independent of $a$. In this case, from \Eq{eq:ndr} it is seen that $\mcc{L}_a$ must be separable as $\mcc{L}_a = \mcc{D}(\omega, \vec{k})A_a$, where $A(a, \omega, \vec{k})$ is some function such that $A_a$ is nonzero. [Parametric dependence of functions like $\mcc{L}$, $\mcc{D}$, and $A$ on $(t,\vec{x})$ is also implied but will be omitted for the sake of brevity.] Then,
\begin{gather}\label{eq:linL0}
\mcc{L} = \mcc{D}(\omega, \vec{k})A.
\end{gather}
It will hence be convenient to think of $a$ as of a \textit{linear} measure of the oscillating field amplitude. Then, most commonly, one will have $A \propto a^2$; yet for our purposes the actual dependence need not be specified.

Equation \eq{eq:ndr} now yields
\begin{gather}\label{eq:linL}
\mcc{D}(\omega, \vec{k}) = 0.
\end{gather}
Thus \Eqs{eq:ii} become
\begin{gather}\label{eq:ii2}
\mc{I} = \mcc{D}_\omega A, \quad \vec{\mc{J}} = - \mcc{D}_\vec{k} A,
\end{gather}
and \Eqs{eq:eqppi} take the form
\begin{gather}\label{eq:eqppi2}
\mc{E} = \omega \mc{I}, \quad \vec{\mc{Q}} = \omega \vec{\mc{J}}, \quad \vec{\mc{P}} = \vec{k}\mc{I}, \quad \oper{\Pi} = \vec{k} \vec{\mc{J}}.
\end{gather}
Hence the photon canonical energy, $H \doteq \mc{E}/\mc{N}$, and the photon canonical momentum, $\vec{P} \doteq \vec{\mc{P}}/\mc{N}$, equal~\cite{ref:sturrock61}
\begin{gather}\label{eq:minkHP}
H = \hbar \omega, \quad \vec{P} = \hbar \vec{k},
\end{gather}
matching the Minkowski interpretation \textit{exactly} and independently of the wave nature. (In fact, $\vec{P} = \hbar \vec{k}$ holds even for nonlinear waves [cf. \Eqs{eq:eqppi}], albeit assuming a fixed ratio of $a$ and the amplitude of each nonnegligible harmonic.) In particular, $P^\alpha \doteq (H/c,\vec{P}) = \hbar k^\alpha$ happens to be a true four-vector, by definition of $k^\alpha$, so $P^\alpha P_\alpha$ is a Lorentz invariant. The latter can also be understood as a measure of the photon canonical mass $\mcc{M}$, defined~via
\begin{gather}\label{eq:meff}
 \mcc{M}^2 \doteq - P^\alpha P_\alpha/c^2
\end{gather} 
(cf., \eg \Refs{ref:kulsrud92, ref:bisnovatyi10, ref:shen04}).

Further, differentiating \Eq{eq:linL} with respect to $\vec{k}$ [with $\omega = \omega(\vec{k}; t, \vec{x})$] also gives ${\mcc{D}_\omega\vec{v}_{\rm g} + \mcc{D}_{\vec{k}} = 0}$, where we introduced the linear group velocity $\vec{v}_{\rm g} \doteq \omega_{\vec{k}}$; therefore,
\begin{gather}
\vec{v}_{\rm g} = - \mcc{D}_\vec{k}/\mcc{D}_\omega = \vec{\mc{J}}/\mc{I}.
\end{gather}
Hence, \Eq{eq:block} yields $\mc{T}^{\alpha\beta} = \mc{N}T^{\alpha\beta}$, where
\begin{gather}\label{eq:block2}
T^{\alpha\beta} =
\left(
\begin{array}{c @{\quad} c}
\hbar\omega & \hbar\omega\vec{v}_{\rm g}/c\\[3pt]
c\hbar\vec{k} & \hbar\vec{k}\vec{v}_{\rm g}
\end{array}
\right)
\end{gather}
is the canonical EMT per photon. Alternatively, one can also exclude $\mc{N}$ and rewrite \Eqs{eq:eqppi2}~as
\begin{gather}\label{eq:linw2}
\vec{\mc{P}} = \vec{k}\mc{E}/\omega, \quad \vec{\mc{Q}} = \mc{E} \vec{v}_{\rm g}, \quad \oper{\Pi} = \vec{\mc{P}} \vec{v}_{\rm g}.
\end{gather}
It is seen, from here and \Eqs{eq:EP}, that the canonical action, energy, and momentum are all transported at the same velocity, $\vec{v}_{\rm g}$. However, keep in mind that the full, or ``kinetic'' energy and momentum densities carried by the wave (\Sec{sec:amc}) generally do not have this property.

Finally, let us introduce photon trajectories, ${d_t \vec{x} = \vec{v}_{\rm g}}$, also known as GO rays. Along those trajectories,
\begin{gather}\label{eq:dt}
d_t = \pd_t + \vec{v}_{\rm g} \cdot \del.
\end{gather}
Then \Eqs{eq:consistvec} yield
\begin{gather}\label{eq:phm}
d_t{\vec{x}} = \vec{v}_{\rm g}, \quad d_t{\vec{k}} = - \omega_{\vec{x}}, \quad d_t{\omega} = \omega_t.
\end{gather}
[Remember that the derivatives $\omega_{\vec{x}}$ and $\omega_t$ of $\omega(\vec{k}; t, \vec{x})$ are taken at fixed $\vec{k}$.] In particular, the ACT can hence be written~as
\begin{gather}\label{eq:lnact}
d_t \ln \mc{I} = - \del \cdot \vec{v}_{\rm g}.
\end{gather}
Also notice that \Eqs{eq:phm} can be understood as canonical equations for the photon motion governed by the Hamiltonian $H(\vec{x}, \vec{P}; t)$. In this form,~\ie
\begin{gather}\label{eq:phm2}
d_t{\vec{x}} = H_\vec{P}, \quad d_t{\vec{P}} = - H_{\vec{x}}, \quad d_t{H} = H_t,
\end{gather}
they are identical to the motion of a true classical particle such as an electron, which supports the well-known analogy between GO and classical mechanics \cite[Sec.~9.8]{book:goldstein}. Reverting to \Eqs{eq:L} and \eq{eq:LL}, it is seen then that not just waves, but classical particles too can be described in terms of phases and amplitudes \cite{foot:meff}.

%--------------------------------------
\subsection{Noether's integrals}
\label{sec:noether}

Various transport equations can now be derived from
\begin{align}
& \pd_t (\msf{X} \mc{I}) + \del \cdot (\msf{X} \vec{\mc{J}}) = \notag\\
& = (\pd_t \msf{X}) \mc{I} + \msf{X} (\pd_t \mc{I}) + (\del \msf{X}) \vec{\mc{J}} + \msf{X} (\del \cdot \vec{\mc{J}}) \notag\\
& = (\pd_t \msf{X}) \mc{I} + (\del \msf{X}) \vec{\mc{J}} \notag\\
& = \mc{I}\,(\pd_t \msf{X} + \vec{v}_{\rm g} \cdot \del \msf{X}) \notag\\
& = \mc{I}\,d_t \msf{X}, \label{eq:aux1}
\end{align}
which holds for arbitrary $\msf{X}$. Some of those are as follows.

\msection{Action} Taking $\msf{X}$ equal to a constant, one recovers \Eq{eq:act2}, or the ACT. [Of course, this is not an independent derivation of the ACT, since the latter itself was used in deriving \Eq{eq:aux1}.] As already emphasized, \Eq{eq:act2} is due to the fact that $\mcc{L}$ does not depend on $\theta$ explicitly. Since it also implies conservation of the integral action $I$, the latter can be understood as the corresponding Noether's integral.

\msection{Energy} Taking $\msf{X} = \omega$, one obtains
\begin{gather}\label{eq:eneq}
\pd_t \mc{E} + \del \cdot (\mc{E}\vec{v}_{\rm g}) = \mc{I}\,d_t \omega.
\end{gather}
As seen from \Eq{eq:phm}, in stationary medium ${d_t \omega = 0}$, so one recovers the result obtained in \Sec{sec:general}, namely, that the wave integral energy, $\int \mc{E}\,dV$, is the Noether's integral that is conserved when the system is translationally invariant in time. Another corollary, which is obtained by comparing \Eq{eq:eneq} with \Eq{eq:e2}, is that 
\begin{gather}\label{eq:eneq2}
-\mcc{L}_t = w = \mc{I}\,d_t \omega = \mc{I}\,\omega_t,
\end{gather}
where we also used \Eq{eq:phm}. Alternatively, one can rewrite this as $w = \mc{N}\,d_t H$, where $d_t H$ is the work on an individual photon per unit time.

\msection{Momentum} Taking $\msf{X} = \vec{k}$, one obtains
\begin{gather}\label{eq:peq}
\pd_t \vec{\mc{P}} + \del \cdot (\vec{\mc{P}}\vec{v}_{\rm g}) = \mc{I}\,d_t \vec{k}.
\end{gather}
As seen from \Eq{eq:phm}, in homogeneous medium ${d_t \vec{k} = 0}$, so one recovers the result obtained in \Sec{sec:general}, namely, that the wave integral momentum, $\int \vec{\mc{P}}\,dV$, is the Noether's integral that is conserved when the system is translationally invariant in space. Another corollary, which is obtained by comparing \Eq{eq:peq} with \Eq{eq:p2}, is that 
\begin{gather}\label{eq:peq2}
\mcc{L}_\vec{x} = \vec{f} = \mc{I}\,d_t \vec{k} = - \mc{I}\,\omega_\vec{x},
\end{gather}
where we also used \Eq{eq:phm}. Alternatively, one can rewrite this as $\vec{f} = \mc{N}\,d_t \vec{P}$, where $d_t \vec{P}$ is the force on an individual photon.

\msection{Angular momentum} Taking $\msf{X} = \vec{x} \times \vec{k}$, one obtains from \Eq{eq:aux1} that
\begin{gather}\label{eq:meq}
\pd_t \vec{\mc{M}} + \del \cdot (\vec{\mc{M}} \vec{v}_{\rm g}) = \mc{I}\,d_t (\vec{x} \times \vec{k}),
\end{gather}
where we formally introduced $\vec{\mc{M}} \doteq (\vec{x} \times \vec{k})\mc{I}$, or
\begin{gather}\label{eq:meqdef}
\vec{\mc{M}} = \vec{x} \times \vec{\mc{P}}.
\end{gather}
Based on \Eq{eq:meqdef}, one could anticipate that $\vec{\mc{M}}$ is the wave angular momentum density, and indeed \Eq{eq:meq} yields that this is the case, as we will now prove.

%--------------------------------------
\subsection{Angular momentum}
\label{sec:angmom}

\msection{Conservation theorem} Consider system rotation by an arbitrary infinitesimal angle $\delta \vec{\varphi}$. Associated with this rotation will be a variation of the Lagrangian density
\begin{gather}\label{eq:dl}
\delta \mcc{L} = \mcc{L}_{\vec{k}} \cdot \delta \vec{k} + \mcc{L}_{\vec{x}} \cdot \delta \vec{x},
\end{gather}
where we substituted \Eq{eq:L} for $\mcc{L}_a$; also, 
\begin{gather}
\delta \vec{k} = \delta \vec{\varphi} \times \vec{k}, \quad \delta \vec{x} = \delta \vec{\varphi} \times \vec{x},
\end{gather}
$\mcc{L}_{\vec{k}} = -\vec{\mc{J}} = - \vec{v}_{\rm g} \mc{I}$, and $\mcc{L}_\vec{x} = \mc{I}\,d_t\vec{k}$, where the latter is taken from \Eq{eq:peq2}. Hence, 
\begin{multline}
\mc{I}^{-1}\delta \mcc{L}
 = - \vec{v}_{\rm g} \cdot (\delta \vec{\varphi} \times \vec{k}) + d_t\vec{k} \cdot (\delta \vec{\varphi} \times \vec{x}) \\
 = \delta \vec{\varphi} \cdot (\vec{v}_{\rm g} \times \vec{k}) + \delta \vec{\varphi} \cdot (\vec{x} \times d_t\vec{k})\\
 = \delta \vec{\varphi} \cdot d_t (\vec{x} \times \vec{k}).
\end{multline}
Having $\delta \mcc{L} = 0$ yields that $d_t (\vec{x} \times \vec{k}) = 0$. From \Eq{eq:meq}, one then obtains that
\begin{gather}
\pd_t \vec{\mc{M}} + \del \cdot (\vec{\mc{M}} \vec{v}_{\rm g}) = 0,
\end{gather}
which means, in particular, that $\int \vec{\mc{M}}\, dV$ is conserved. Since this is the invariant associated with the medium isotropy, it by definition \cite[Sec.~9]{book:landau1} represents the wave angular momentum. Correspondingly, $\vec{\mc{M}}$ is the wave angular momentum density \cite{foot:mdef}. Also, $\vec{M} \doteq \vec{\mc{M}}/\mc{N}$, or
\begin{gather}
\vec{M} = \vec{x} \times \vec{P},
\end{gather}
is the angular momentum of a photon, $\hbar\, {d_t (\vec{x} \times\vec{k})} \equiv d_t\vec{M}$ is the torque on a photon (cf. \Ref{ref:jones73}), and the corresponding dynamic equation is spelled out as
\begin{gather}\label{eq:mtor}
d_t \vec{M} = \vec{v}_{\rm g} \times \vec{P} - \vec{x} \times H_\vec{x}.
\end{gather}

\msection{Spin angular momentum (SAM)} Consider a stationary wave beam symmetric with respect to $z$ axis; \ie in cylindrical coordinates $(r, \phi, z)$, the amplitude $a$ and the wave vector components $k_r$, $k_\phi$, $k_z$ are independent of $\phi$. The consistency relation \eq{eq:consistvec} requires then that $\pd_r (r k_\phi) = 0$, so $k_\phi = m/r$, where $m$ is a constant. This gives $\mc{M}_z = rk_\phi \mc{I} = m\mc{I}$, or that the carried angular momentum per photon is $M_z = m\hbar$. To find $m$, notice that, due to $k_\phi = r^{-1}\pd_\phi \theta$, the wave canonical phase has the form $\theta = m \phi - \omega t + \Xi(r,z)$, where $\Xi$ is some function of $r$ and $z$ only. Thus, after any time $\delta t$, the wave must repeat itself, at the same $r$ and $z$, in the coordinate frame rotated by $\delta \phi = (\omega/m)\,\delta t$. Satisfying this condition are, in fact, only circularly polarized waves (at least, in free space), corresponding to $m = \pm 1$. Other types of wave beams therefore cannot be considered symmetric within GO and thus can be assigned only average $m$. Specifically, decomposing a wave with a given elliptic polarization into the two independent circularly-polarized components with corresponding weights $C_+$ and $C_-$, one gets $\favr{m} = C_+ - C_-$. In particular, linear polarization corresponds to $C_+ = C_-$, in which case ${\favr{m} = 0}$.

These results match the known quantum theorem, which says that states with circular polarization are the only polarization states of a free photon that are eigenstates of the corresponding SAM projection, $M_z = \pm \hbar$~\cite[Sec.~8]{book:landau4}. Thus, for an axially symmetric beam, $\mc{M}_z$ that originates entirely from the beam polarization can be called the SAM density. Interestingly, it can also be interpreted as follows. For those (circularly polarized) waves that do allow precise definition of the SAM, the latter appears due to the singularity of $k_\phi$ at $r = 0$, \ie due to $\theta(r = 0)$ being undefined \cite{foot:singo}. In this sense, the canonical phase increment $\Delta\theta = 2\pi m$ along a closed contour encircling the symmetry axis is the corresponding Berry phase \cite{ref:hannay85, ref:berry84}, so the photon SAM (in units~$\hbar$) is nothing but the Berry index of the classical phase field. 

Finally, note that a wave beam that is not axially symmetric will also carry additional, ``orbital'' momentum~\cite{ref:allen92, tex:allen99}. The latter is included in \Eq{eq:meqdef}, and separating it from the SAM unambiguously may not be possible except in special cases, as usual; see, \eg \Ref{ref:barnett94, ref:allen92, tex:allen99} or Ref.~\cite[Sec.~6]{book:landau4}.

%--------------------------------------
\subsection{Dissipation}
\label{sec:dissw}

Suppose now that a linear wave experiences weak dissipation. Then, comprising the wave locally are Fourier harmonics with \textit{complex} frequencies and wave vectors,
\begin{gather}
\Omega = \Omega' + i \Omega'', \quad \vec{K} = \vec{K}' + i \vec{K}''.
\end{gather}
Assuming the local dispersion relation in the form
\begin{gather}\label{eq:dr2}
\mcc{D}(\Omega, \vec{K}) = 0,
\end{gather}
let us keep only the terms of the zeroth and first order in $\Omega''$ and $\vec{K}''$. Then one gets
\begin{gather}\label{eq:dr3}
\mcc{D} + i \mcc{D}_\Omega \Omega'' + i \mcc{D}_\vec{K} \cdot \vec{K}'' = 0,
\end{gather}
where $\mcc{D}$ and its derivatives are henceforth evaluated at $(\Omega', \vec{K}')$. Now suppose $\mcc{D} = \mcc{D}'+ i \mcc{D}''$, where $\mcc{D}'' \doteq \text{Im}\,\mcc{D}$ is much smaller than $\mcc{D}' \doteq \text{Re}\,\mcc{D}$. One hereby obtains
\begin{gather}\label{eq:dr5}
\mcc{D}' + i \mcc{D}'' + i \mcc{D}'_\Omega \Omega'' + i \mcc{D}'_\vec{K} \cdot \vec{K}'' = 0
\end{gather}
(where higher-order terms were neglected), the real part and the imaginary part of which are, correspondingly,
\begin{gather}
\mcc{D}' = 0,\label{eq:dr5r}\\
\mcc{D}'' + \mcc{D}'_\Omega \Omega'' + \mcc{D}'_\vec{K} \cdot \vec{K}'' = 0.\label{eq:dr5i}
\end{gather}
From here, the envelope dynamics is inferred as follows.

At any given time, the field distribution of the real system can be mapped into the auxiliary nondissipative system, where the wave phase $\theta$ is well defined, and
\begin{gather}
\mcc{L} \doteq \mcc{D}'(\omega, \vec{k})A.
\end{gather}
This defines the instantaneous $a$ and also the instantaneous \textit{real} canonical frequency and wave vector, $(\omega, \vec{k})$; hence all other local quantities can be introduced through $\mcc{L}(a, \omega, \vec{k})$ too. However, the dynamics in the auxiliary system and in the real system are different; thus, for the latter, an extra term $\Gamma$ must be added in \Eq{eq:lnact},
\begin{gather}\label{eq:lnact2}
d_t \ln \mc{I} = - \del \cdot \vec{v}_{\rm g} - \Gamma.
\end{gather}

Assume that dissipation is determined by the local $(a, \omega, \vec{k})$ and by the local parameters of the medium, rather than their gradients. Then one can find $\Gamma$ by calculating it for homogeneous stationary medium and a wave whose field is locally ``monochromatic'', \ie can be assigned particular complex $(\Omega, \vec{K})$ [which map to the given canonical $(\omega,\vec{k})$]. Then,
\begin{gather}\label{eq:aux3}
\Gamma = - d_t \ln \mc{I} = - \varkappa(\Omega'' - \vec{v}_{\rm g} \cdot \vec{K}''),
\end{gather}
where $\varkappa \doteq d\ln A/d\ln a$ (which commonly equals~2; see \Sec{sec:linbasic}), and the left-hand side is evaluated at $(\Omega', \vec{K}')$. On the other hand, \Eq{eq:dr5i} yields
\begin{gather}\label{eq:dr6}
\Omega'' - \vec{v}_{\rm g} \cdot \vec{K}'' = - \mcc{D}''/\mcc{D}'_\Omega.
\end{gather}
Hence $\Gamma$ is connected with the dispersion function as
\begin{gather}\label{eq:Gamma}
\Gamma(\omega, \vec{k}) = \varkappa\,\mcc{D}''(\omega, \vec{k})/\mcc{D}'_\omega(\omega, \vec{k}),
\end{gather}
where we used that, to the leading order, it is sufficient to take $(\Omega', \vec{K}') \approx (\omega, \vec{k})$ on the right-hand side.

Now let us present the corresponding transport equations. Similarly to \Eq{eq:aux1}, one has, for any $\msf{X}$, that
\begin{gather}
\pd_t (\msf{X} \mc{I}) + \del \cdot (\msf{X} \vec{\mc{J}}) = \mc{I}\,d_t \msf{X} - \Gamma \msf{X} \mc{I}. \label{eq:aux11}
\end{gather}
Since $\msf{X}$ is arbitrary, the number of equations that can be produced from here is infinite, like in \Sec{sec:noether}. In particular, those for the action, the energy, the momentum, and the angular momentum are obtained by taking $\msf{X} = 1$, $\msf{X} = \omega$, $\msf{X} = \vec{k}$, and $\msf{X} = \vec{x} \times \vec{k}$, correspondingly, and are as follows:
\begin{gather}
\pd_t \mc{I} + \del \cdot (\mc{I}\vec{v}_{\rm g}) = - \Gamma \mc{I}, \label{eq:ae}\\
\pd_t \mc{E} + \del \cdot (\mc{E}\vec{v}_{\rm g}) = \mc{I}\,d_t\omega - \Gamma \mc{E}, \\
\pd_t \vec{\mc{P}} + \del \cdot (\vec{\mc{P}} \vec{v}_{\rm g}) = \mc{I}\, d_t \vec{k} - \Gamma \vec{\mc{P}}, \\
\pd_t \vec{\mc{M}} + \del \cdot (\vec{\mc{M}} \vec{v}_{\rm g}) = \mc{I}\, d_t (\vec{x} \times \vec{k}) - \Gamma \vec{\mc{M}}. \label{eq:me}
\end{gather}
The physical statement contained in these is twofold. First of all, one can see that the decay rate is the same in all the equations, regardless of the specific $\msf{X}$. [This, of course, is seen already from \Eq{eq:aux11}.] Second of all, this rate is actually \textit{known} from \Eq{eq:Gamma}, which connects $\Gamma$ with the dispersion function $\mcc{D}$. In particular, the action loss per unit volume per unit time can be written~as
\begin{gather}
\imath_{\rm loss} \doteq \Gamma \mc{I} = \varkappa\,\mcc{D}'' A, \label{eq:id1}
\end{gather}
and the corresponding losses of the wave energy, momentum, and angular momentum are given by
\begin{gather}
w_{\rm loss} = \omega \imath_{\rm loss}, \quad 
\vec{f}_{\rm loss} = \vec{k} \imath_{\rm loss}, \quad
\vec{\kappa}_{\rm loss} = (\vec{x}\times \vec{k}) \imath_{\rm loss}.\notag
\end{gather}
Also notice that $d_t\omega$ and $d_t \vec{k}$ entering \Eqs{eq:ae}-\eq{eq:me} can be taken from the GO ray equations. Since based entirely on \Eqs{eq:omk} and \eq{eq:consistvec} (\Sec{sec:linbasic}), those happen to be unaffected by dissipation; \ie they are still given by \Eqs{eq:phm}. Hence, the above results can be interpreted as follows: local dissipation does not affect individual photons but rather changes the photon density.

For an explanation of how the results reported here apply to electromagnetic waves, see \Sec{sec:em}. The same results are also applicable to dissipation-driven instabilities ($\Gamma < 0$). Nondissipative instabilities can be accommodated within GO too, namely, by allowing for complex rays; for details see \Ref{ref:startsev09} and references therein.

%%%%%%%%%%%%%%%%%%%%%%%%%%%%%%%%%%%%%%%
\section{Linear waves: Abraham representation}
\label{sec:amc}

%--------------------------------------
\subsection{Basic definitions} 

In addition to the wave canonical, or Minkowski EMT that we discussed so far, one can also introduce the corresponding so-called kinetic, or Abraham EMT,
\begin{gather}\label{eq:kintau}
\tau^{\alpha\beta} = \left(
\begin{array}{c @{\quad} c}
\varepsilon & \vec{\vartheta}/c\\[3pt]
c\vec{\rho} & \oper{\pi}
\end{array}
\right).
\end{gather}
It is defined such that, being a part of the complete EMT that describes the ``wave + medium'' system (WMS), $\tau^{\alpha\beta}$ comprises all the wave-related (\ie $a$-dependent) dynamics of the medium and fields. We hence express it as $\tau^{\alpha\beta} = \mc{T}^{\alpha\beta} + \Delta\tau^{\alpha\beta}$, where $\Delta\tau^{\alpha\beta}$ is the ``ponderomotive'' part that is stored in the medium, and, similarly,
\begin{gather}\label{eq:fullpm}
\varepsilon = \mc{E} + \Delta\varepsilon, \quad
\vec{\rho} = \vec{\mc{P}} + \Delta\vec{\rho}, \quad
\vec{\mu} = \vec{\mc{M}} + \Delta\vec{\mu}.
\end{gather}

In particular, notice the following. Since the WMS is closed and thus Lorentz-invariant, its complete EMT is symmetrizable \cite{ref:belinfante39, ref:belinfante40, ref:dewar77}. Yet its unperturbed part is symmetrizable by itself (because it describes a closed system too, namely, the medium absent a wave), so $\tau^{\alpha\beta}$ is also symmetrizable separately. On the other hand, since $\tau^{\alpha\beta}$ is proportional to the wave intensity, it is defined uniquely and, therefore, \textit{must} be symmetric. This yields $\vec{\rho} = \vec{\vartheta}/c^2$, and
\begin{gather}\label{eq:mufin}
\vec{\mu} = \vec{x} \times \vec{\rho}
\end{gather}
holds automatically \cite[Sec.~32]{book:landau2}. Also, since the \textit{integral} energy-momentum of the whole WMS is defined uniquely \cite[Sec.~32]{book:landau2}, and its $a$-dependent part is defined uniquely too, one can find $(\varepsilon/c,\vec{\rho})$ as the $a$-dependent part of the WMS \textit{canonical} energy-momentum density. Given the WMS Lagrangian density, the latter can, in principle, be found straightforwardly in any specific problem \cite{foot:noether}. However, the general answer is not informative (meaning that $\tau^{\alpha\beta}$ is by itself a somewhat artificial construct; see also \Ref{book:hehl}). Thus, below, we consider only the particular model of an isotropic medium, most popular in the AMC context, yet still refrain from specifying the wave nature.

%--------------------------------------
\subsection{Wave energy-momentum in isotropic medium}

\msection{General case} Consider an isotropic medium (such as gas, fluid, or plasma) comprised of elementary \cite{foot:elem} particles or fluid elements whose dynamics absent a wave is described by some aggregate Lagrangian $L$. In the presence of a wave, the WMS Lagrangian is hence $L + \msf{L}$, where $\msf{L} = \int \mcc{L}\,dV$ is the wave Lagrangian. Assuming that particles contribute to $\msf{L}$ additively, the latter can be written as $\msf{L} = \msf{L}^{(0)} - \sum_\ell \Phi^{(\ell)}$, where $\msf{L}^{(0)}$ is independent of all particle velocities $\vec{u}^{(\ell)}$, and each of the so-called ponderomotive potentials $\Phi^{(\ell)}$ \cite{my:bgk, my:acti}, or dipole potentials \cite{my:nlinphi, my:kchi}, depends on the specific $\vec{u}^{(\ell)}$ but not on other velocities. Omitting the index $\ell$, we can write the canonical momentum of each particle as the sum of the mechanical part $\pd_\vec{u} L$ and the ponderomotive part $- \pd_\vec{u} \Phi$, also yielding the ponderomotive contribution to the particle canonical energy, $- \vec{u} \cdot \pd_\vec{u}\Phi$. (This energy should not be confused with the ponderomotive potential $\Phi$ itself, which is a part of the \textit{wave} canonical energy \cite{foot:kchi}.) Thus, the densities of the ponderomotive momentum and energy stored in particles can be written as follows:
\begin{gather}
\Delta\vec{\rho} = - \sum_s n^{(s)}\favr{\pd_\vec{u} \Phi}^{(s)},\label{eq:dpdef}\\
\Delta\varepsilon = - \sum_s n^{(s)}\favr{\vec{u} \cdot \pd_\vec{u} \Phi}^{(s)},\label{eq:dedef}
\end{gather}
where the summation is taken over different species, $n^{(s)}$ are the (locally averaged) densities of those species, and angular brackets denote averaging over velocities within the corresponding ensembles.

\msection{Fluid model} If a medium can be modeled as a single fluid (in particular meaning that kinetic effects are inessential, unlike, \eg in warm plasma), one can simplify \Eqs{eq:dedef} and \eq{eq:dpdef} further, namely, as follows. First of all, notice that the velocities $\vec{u}$ of fluid elements are all equal to a single velocity $\vec{v}$, so \Eqs{eq:dpdef} and \eq{eq:dedef} become
\begin{gather}\label{eq:drde}
\Delta\vec{\rho} = - n\,\pd_\vec{v}\Phi, \quad \Delta\varepsilon = \vec{v} \cdot \Delta\vec{\rho}.
\end{gather}
It is hence convenient to rewrite \Eqs{eq:drde} in terms of Lorentz-invariant proper parameters of the medium \cite{foot:proper}. Since $\Phi$ that enters here depends on the wave intensity, it must be gauge-invariant; thus, being (minus) the interaction Lagrangian of a single element, it transforms as $\Phi = \Phi'/\gamma$ \cite{my:mneg}, with primes \textit{in this section} (\Sec{sec:amc}) denoting the medium rest frame, and $\gamma = (1 - v^2/c^2)^{-1/2}$. Also, $n = \gamma n'$, where $n'$ is the proper density, correspondingly. Since the latter does not depend on $\vec{v}$, we then get $\Delta\vec{\rho} = - \pd_\vec{v}(n'\Phi') + \gamma^2\vec{v}n'\Phi'/c^2$. Further, let us denote
\begin{gather}\label{eq:mcUdef}
n'\Phi' = \mcc{L}' - \mcc{L}'^{(0)} \doteq \mc{U}',
\end{gather}
where $\mcc{L}'^{(0)}$ is $\msf{L}'^{(0)}$ per unit volume, and introduce
\begin{gather}
\vec{\mcc{R}} \doteq \frac{\gamma^2\vec{v}}{c^2}\,\mc{U}', \label{eq:mccR}
\end{gather}
understood as the striction contribution (\Sec{sec:kin}). Since $\mcc{L}'^{(0)}$ is also independent of~$\vec{v}$, one then can write
\begin{gather}\label{eq:aux59}
\Delta\vec{\rho} = \pd_\vec{v}\mcc{L}' + \vec{\mcc{R}}.
\end{gather}

Due to the fact that a Lagrangian density is a four-scalar, $\mcc{L}'$ that enters \Eq{eq:aux59} can also be replaced with $\mcc{L}$ \cite{foot:flagr}. However, using $\mcc{L}'(a', k'_\mu)$ is preferable, because it cannot depend on $\vec{v}$ explicitly, but rather depends on it solely through $a'$ and $k'_\mu$. [Remember that the velocity derivative in \Eq{eq:aux59} must be taken at fixed $a$ and $k_\mu$.] Due to $\mcc{L}'_{a'} = 0$ [cf. \Eq{eq:linL}], we then get
\begin{gather}\label{eq:aux50}
\pd_\vec{v}\mcc{L}' = - (\pd_\vec{v}{\Lambda^\nu}_\mu)\,k_\nu \mc{J}'^\mu,
\end{gather}
where we substituted the (covector) Lorentz transformation \eq{eq:lorentz}, \ie $k'_\mu = {\Lambda^\nu}_\mu k_\nu$. On the other hand, $k_\nu = {(\Lambda^{-1})^\lambda}_\nu k'_\lambda$, so \Eq{eq:aux50} can also be written as
\begin{gather}\label{eq:aux57}
\pd_\vec{v}\mcc{L}' = - \gamma{\vec{G}^\lambda}_\mu {\mc{T}'_\lambda}^\mu/c,
\end{gather}
where we introduced a dimensionless matrix function
\begin{gather}\label{eq:gbardef}
{\vec{G}^\lambda}_\mu(\vec{v}) \doteq (c/\gamma){(\Lambda^{-1})^\lambda}_\nu\, (\pd_\vec{v}{\Lambda^\nu}_\mu).
\end{gather}
As shown in Appendix, \Eq{eq:aux57} is also equivalent to 
\begin{gather}\label{eq:aux58}
\pd_\vec{v} \mcc{L}' = \gamma\text{Tr}(\vec{G}\mc{T}')/c = \vec{\mcc{P}} + \vec{\mcc{B}},
\end{gather}
where the terms on the right-hand side are defined as
\begin{gather}
\vec{\mcc{P}} = \gamma\oper{\Lambda} \cdot \left(\frac{\mc{E}' \vec{v}'_{\rm g}}{c^2} - \vec{\mc{P}}'\right),\label{eq:mccP}\\
\vec{\mcc{B}} = \frac{\gamma^2}{\gamma + 1}\left[\frac{\vec{v}}{c} \times \left(\frac{\vec{v}'_{\rm g}}{c} \times \vec{\mc{P}}'\right)\right].\label{eq:mccB}
\end{gather}
Yet, $\vec{v}_{\rm g}'$ is parallel to $\vec{k}'$ in isotropic medium, so $\vec{\mcc{B}}$ vanishes, and we finally get
\begin{gather}\label{eq:DeltaRho}
\Delta\vec{\rho} = \vec{\mcc{P}} + \vec{\mcc{R}}, \quad
\Delta\varepsilon = \vec{v} \cdot (\vec{\mcc{P}} + \vec{\mcc{R}}).
\end{gather}

%--------------------------------------
\subsection{Wave EMT in the isotropic-fluid model}
\label{eq:kinemtfl}

Within the isotropic-fluid model, one can hence explicitly construct the complete kinetic EMT of a wave,
\begin{gather}\label{eq:trans}
\tau^{\alpha\beta} = {\Lambda^\alpha}_\mu {\Lambda^\beta}_\nu \tau'^{\mu\nu},
\end{gather}
which is done as follows. 

\msection{Energy and momentum} First of all, let us combine \Eqs{eq:fullpm} and \eq{eq:DeltaRho} with \Eq{eq:mccP} for $\vec{\mcc{P}}$, \Eq{eq:mccR} for $\vec{\mcc{R}}$, and $\mc{E} = \omega \mc{I}$ and $\vec{\mc{P}} = \vec{k} \mc{I}$, as well as with
\begin{gather}
\mc{I} = \gamma \mc{I}'(1 + \vec{v} \cdot \vec{v}_{\rm g}'/c^2),
\end{gather}
where we employed the four-vector transformation properties of $\mc{J}^\alpha$. This yields
\begin{gather}
\varepsilon = \gamma^2 \mc{E}' + \frac{\gamma \mc{E}' \vec{v}}{c^2} \cdot \left(\oper{\Lambda} \cdot \vec{v}_{\rm g}' + \frac{\omega}{\omega'}\,\vec{v}_{\rm g}'\right) + 
\frac{\gamma^2v^2}{c^2}\, \mc{U}',\label{eq:vefin}\\
\vec{\rho} = 
\frac{\gamma \mc{E}'}{c^2}\left[\oper{\Lambda} \cdot \vec{v}_{\rm g}' + \gamma\vec{v} + \frac{\vec{k}}{\omega'}(\vec{v} \cdot \vec{v}_{\rm g}')\right]
+ \frac{\gamma^2\vec{v}}{c^2}\, \mc{U}'.\label{eq:pfin}
\end{gather}
[Entering the numerator in \Eq{eq:pfin} is actually $\vec{k}$, not $\vec{k}'$.] By taking $\vec{v} = 0$ here, we then get, in particular,
\begin{gather}
\varepsilon' = \mc{E}', \quad c\vec{\rho}' = \vec{\vartheta}'/c = \mc{E}'\vec{v}'_{\rm g}/c,\label{eq:primedep}
\end{gather}
also using that $\tau^{\alpha\beta}$ is symmetric in all reference frames.

\msection{Momentum flux density} Since $\vec{k}'$ is the only designated direction in the medium rest frame, the (symmetric) momentum flux density $\oper{\pi}'$ must be a linear superposition of $\vec{k}'\vec{k}'$ and $\oper{1}'$, or, equivalently, $\oper{\pi}' = \psi\,\vec{k}'\vec{v}_{\rm g}' + \zeta\, \oper{1}'$, where $\psi$ and $\zeta$ are some coefficients. Combining this with \Eqs{eq:trans} and \eq{eq:primedep} and plus with, \eg \Eq{eq:vefin} for $\varepsilon \equiv \tau^{00}$, one readily obtains $\psi = \mc{I}'$ and $\zeta = \mc{U}'$; \ie
\begin{gather}
\oper{\pi}' = \mc{E}'\,\vec{k}'\vec{v}_{\rm g}'/\omega' + \mc{U}'\,\oper{1}'.
\end{gather}
(In particular, if $\vec{v}_{\rm g}' = 0$, the term $\mc{U}'$ acts as the ponderomotive pressure; cf. \Ref{my:dense}.) Equation \eq{eq:pfin} then flows from \Eq{eq:trans} automatically; yet, \Eq{eq:trans} also gives
\begin{multline}\label{eq:sigfin}
\oper{\pi} = 
\frac{\omega'(\vec{k}'\cdot \vec{v}_{\rm g}'\mc{E}')}{c^2|\vec{k}'|^2}\left(\frac{c^2\vec{k}\vec{k}}{\omega'^2} - \frac{\gamma^2\vec{v}\vec{v}}{c^2}\right)\\
+ \frac{\gamma\vec{v}\vec{v}}{c^2}\,\mc{E}' + \left(\oper{1} + \frac{\gamma^2\vec{v}\vec{v}}{c^2}\right)\mc{U}'.
\end{multline}

\msection{EMT and ponderomotive forces} The wave kinetic EMT in isotropic fluid is hereby summarized as
\begin{gather}
\tau'^{\alpha\beta} = 
\left(
\begin{array}{c @{\quad} c}
\mc{E} & \mc{E}\vec{v}_{\rm g}/c\\[3pt]
\mc{E}\vec{v}_{\rm g}/c & \mc{E}\,\vec{k}\vec{v}_{\rm g}/\omega + \mc{U}\,\oper{1}
\end{array}
\right)'
\end{gather}
in the medium rest frame and is transformed to other frames via \Eq{eq:trans}, as also spelled out in \Eqs{eq:vefin}, \eq{eq:pfin}, and \eq{eq:sigfin}. In particular, if the flow velocity is negligible in a given frame, one can take ${\Lambda^\alpha}_\beta \approx \delta^\alpha_\beta$, so
\begin{gather}\label{eq:pqcex}
\varepsilon \approx \mc{E}, \quad \vec{\rho} \approx \mc{E} \vec{v}_{\rm g}/c^2, \quad \vec{\mu} \approx (\vec{x} \times \vec{v}_{\rm g}) \mc{E}/c^2.
\end{gather}
Finally, the ponderomotive four-force density $\bar{f}^\alpha$ that a wave imparts to a medium also can be calculated \cite{foot:sign},
\begin{gather}
\bar{f}^\alpha = - {\tau^{\alpha\beta}}_{;\beta},
\end{gather}
whence, substituting $\bar{f}^\alpha = (\bar{w}/c, \bar{\vec{f}})$, one obtains
\begin{gather}
\bar{w} = - \pd_t \varepsilon - c^2 \del \cdot \vec{\rho}, \quad
\bar{\vec{f}} = - \pd_t \vec{\rho} - \del \cdot \oper{\pi}.
\end{gather}
(Here $\bar{w}$ has the meaning of the power density input into the medium, and $\bar{\vec{f}}$ is the usual three-force density.) In particular, note that since $\tau^{\alpha\beta}$ is expressed through quantities derived from $\mcc{L}'$ and $\mcc{L}'^{(0)}$, which are the fundamental invariants of the wave, the usual ambiguity in calculating the forces on the medium is hence avoided.

The above results, which rely essentially \textit{only} on \Eq{eq:L} and the isotropic-fluid approximation (without any reference to electromagnetism), represent a more concise and transparent version of those reported in \Ref{ref:dewar77} and generalize the latter to the case of waves of arbitrary nature; see also \Sec{sec:kin}.

%--------------------------------------
\subsection{Photon kinetic properties}
\label{sec:abr}

The following energy, momentum, and angular momentum can now be assigned to a single photon:
\begin{gather}\label{eq:pmdefkin}
h \doteq \varepsilon/\mc{N}, \quad \vec{p} \doteq \vec{\rho}/\mc{N}, \quad \vec{m} \doteq \vec{\mu}/\mc{N}.
\end{gather}
Keep in mind, however, that these are merely quantities \textit{per}~photon rather than the momenta \textit{of}~a photon, in contrast with $(H, \vec{P}, \vec{M})$ that actually enter the photon motion equations [\Eqs{eq:phm2} and \eq{eq:mtor}]. As a result, $(h, \vec{p}, \vec{m})$ do not enjoy the simple transformation properties of their canonical counterparts. In particular, the kinetic four-momentum $p^\alpha \doteq (h/c, \vec{p})$ is generally not a four-vector. One can easily check this, \eg by using $p^\alpha = (cN)^{-1}\int \tau^{\alpha0}\,dV$ with $\tau^{\alpha\beta}$ taken from \Sec{eq:kinemtfl} and
\begin{gather}
\mc{N} = \gamma \mc{N}'(1 + \vec{v} \cdot \vec{v}_{\rm g}'/c^2),\\ \mc{E}' = \hbar\omega'\mc{N}', \quad \vec{\mc{P}}' = \hbar \vec{k}'\mc{N}'.\label{eq:EPprimed}
\end{gather}

Still, simple expressions are obtained from \Eqs{eq:pqcex} for isotropic fluid medium at rest; namely,
\begin{gather}\label{eq:abrappr}
h \approx \hbar \omega, \quad \vec{p} \approx \hbar\omega\vec{v}_{\rm g}/c^2, \quad \vec{m} \approx (\vec{x} \times \vec{v}_{\rm g}) \hbar\omega/c^2.
\end{gather}
Since here $\vec{v}_{\rm g}$ is assumed to be parallel to $\vec{k}$, one also gets that $\vec{p}$ is parallel to $\vec{P}$, $\vec{m}$ is parallel to $\vec{M}$, and
\begin{gather}
p/P = m/M \approx 1/(n_{\rm p} n_{\rm g}).
\end{gather}
These match the traditional Abraham's formulas \cite{ref:barnett10, ref:padgett03}, hence seen to hold for waves of arbitrary (not necessarily electromagnetic) nature. Yet it is clear now that the traditional formulas are, in fact, approximate and generally invalid for moving and non-fluid media, in contrast with Minkowski's formulas for the canonical quantities [\Eqs{eq:minkHP} and \eq{eq:meqdef}], which are more universal.

%%%%%%%%%%%%%%%%%%%%%%%%%%%%%%%%%%%%%%%
\section{Linear electromagnetic waves}
\label{sec:em}

Finally, let us apply the above results to illustrate how the properties of linear electromagnetic waves can be calculated explicitly within our general approach, without using Maxwell's equations for the wave envelope. Note also that similar calculations can be performed for nonlinear waves too, for which $\mcc{L}$ can be constructed from first principles as well \cite{my:bgk, my:acti, my:actii, my:actiii}.

%--------------------------------------
\subsection{Wave Lagrangian}
\label{sec:U}

\msection{Basic equations} First, let us consider a nondissipative wave, as usual. The wave Lagrangian density (derived independently, \eg in \Refs{ref:dewar77, ref:philbin11}) can be expected in the form $\mcc{L} = \mcc{L}^{(0)} - \mc{U}$, where
\begin{gather}\label{eq:l0}
\mcc{L}^{(0)} \doteq \frac{1}{16\pi}\,(\tilde{\vec{E}}^* \cdot \tilde{\vec{E}} - \tilde{\vec{B}}^* \cdot \tilde{\vec{B}})
\end{gather}
is that in vacuum \cite{my:bgk}, $\tilde{\vec{E}}$ and $\tilde{\vec{B}}$ are the electric and magnetic field envelopes, and $\mc{U}$ is the potential energy density of the wave-medium interaction [cf. \Eq{eq:mcUdef}]. For linear, \ie dipolar interaction, we can take \cite[Secs.~4.2, 4.8, 5.7, 6.2]{book:jackson}
\begin{gather}\label{eq:lint3}
\mc{U} = - \frac{1}{4}\,\text{Re}\,\left(\tilde{\vec{E}}^* \cdot \tilde{\vec{P}} + \tilde{\vec{B}}^* \cdot \tilde{\vec{M}}\right).
\end{gather}
Here $\tilde{\vec{P}}$ is the electric dipole moment density (\ie the polarization), and $\tilde{\vec{M}}$ is the magnetic dipole moment density (\ie the magnetization); also, one factor $1/2$ comes from the time-averaging, and the other $1/2$ comes from the fact that $\tilde{\vec{P}}$ and $\tilde{\vec{M}}$ are linear functions of $\tilde{\vec{E}}$ and $\tilde{\vec{B}}$, correspondingly. Now let us introduce $\tilde{\vec{D}}$ and $\tilde{\vec{H}}$ via
\begin{gather}
\tilde{\vec{D}} \doteq \tilde{\vec{E}} + 4\pi \tilde{\vec{P}} \doteq \oper{\epsilon}\cdot\tilde{\vec{E}}, \\
\tilde{\vec{B}} \doteq \tilde{\vec{H}} + 4\pi \tilde{\vec{M}} \doteq \oper{\mu} \cdot\tilde{\vec{H}},
\end{gather}
assuming that the permittivity tensor $\oper{\epsilon}$ and the permeability tensor $\oper{\mu}$ (not to be confused with the kinetic angular momentum density $\vec{\mu}$) are Hermitian so the assumption of zero dissipation be satisfied. One gets then~\cite{ref:dewar77, ref:philbin11}
\begin{gather}
\mcc{L} =\frac{1}{16\pi}\,\left(
\tilde{\vec{E}}^* \cdot \oper{\epsilon} \cdot \tilde{\vec{E}} - 
\tilde{\vec{B}}^* \cdot \oper{\mu}^{-1} \cdot \tilde{\vec{B}}
\right)
\label{eq:lem}
\end{gather}
(here $\oper{\mu}^{-1}$ is the tensor inverse to $\oper{\mu}$), also meaning that
\begin{gather}\label{eq:mcUel}
\mc{U} = - \frac{1}{16\pi}\,\left[
\tilde{\vec{E}}^* \cdot (\oper{\epsilon} - \oper{1}) \cdot \tilde{\vec{E}} - 
\tilde{\vec{B}}^* \cdot (\oper{\mu}^{-1} - \oper{1}) \cdot \tilde{\vec{B}}
\right].
\end{gather}
In agreement with \Refs{my:bgk, my:acti}, this implies assigning the following ponderomotive potentials to particles (or fluid elements) comprising the medium:
\begin{gather}\label{eq:Phia}
\Phi = - \tilde{\vec{E}}^*\cdot\oper{\alpha}\cdot\tilde{\vec{E}}/4 - \tilde{\vec{B}}^*\cdot\oper{\beta}\cdot\tilde{\vec{B}}/4,
\end{gather}
where $\oper{\alpha}$ and $\oper{\beta}$ are the particle electric and magnetic polarizabilities \cite{my:kchi}, and
\begin{gather}
\oper{\epsilon} = \oper{1} + \sum_s 4\pi n^{(s)} \favr{\oper{\alpha}}^{(s)}, \label{eq:epsfl}\\
\oper{\mu}^{-1} = \oper{1} - \sum_s 4\pi n^{(s)} \favr{\oper{\beta}}^{(s)}. \label{eq:mufl}
\end{gather}

\msection{Parametrization and dispersion} Remember that there is a freedom in defining $a$, so there are various options for how to parametrize the wave Lagrangian density. First, let us consider $\tilde{\vec{E}}$ and $\tilde{\vec{E}}^*$ as independent vector fields; \ie $a = (\tilde{\vec{E}},\tilde{\vec{E}}^*)$. In this case, it is convenient to write
\begin{gather}\label{eq:l02}
\mcc{L}^{(0)} = \frac{1}{16\pi} \left(\tilde{\vec{E}}^* \cdot \tilde{\vec{E}} - \frac{c^2}{\omega^2}\,|\vec{k} \times \tilde{\vec{E}}|^2\right)
\end{gather}
(where we used that $\tilde{\vec{B}} = c\vec{k} \times \tilde{\vec{E}}/\omega$) and
\begin{gather}\label{eq:lagr1}
\mcc{L} = \frac{1}{16\pi} \left[
\tilde{\vec{E}}^* \cdot \oper{\epsilon} \cdot \tilde{\vec{E}} - 
\frac{c^2}{\omega^2}\,(\vec{k} \times \tilde{\vec{E}}^*) \cdot \oper{\mu}^{-1} \cdot (\vec{k} \times \tilde{\vec{E}})
\right],
\end{gather}
correspondingly. Using that
\begin{multline}\label{eq:ke}
(\vec{k} \times \tilde{\vec{E}}^*) \cdot \oper{\mu}^{-1} \cdot (\vec{k} \times \tilde{\vec{E}}) = \\
- (\tilde{\vec{E}}^* \times \vec{k}) \cdot \oper{\mu}^{-1} \cdot (\vec{k} \times \tilde{\vec{E}}) = \\
- \tilde{\vec{E}}^* \cdot \{\vec{k} \times [\oper{\mu}^{-1} \cdot (\vec{k} \times \tilde{\vec{E}})]\},
\end{multline}
one can further rewrite \Eq{eq:lagr1} as follows:
\begin{gather}
\mcc{L} = \frac{\tilde{\vec{E}}^*}{16\pi}\cdot\bigg\{
\oper{\epsilon} \cdot \tilde{\vec{E}} + \frac{c^2}{\omega^2}\,
\vec{k} \times \Big[\oper{\mu}^{-1} \cdot (\vec{k} \times \tilde{\vec{E}})\Big]
\bigg\}.
\end{gather}
Then, varying $\mcc{L}$ with respect to $\tilde{\vec{E}}^*$ yields the following dispersion relation:
\begin{gather}\label{eq:drem}
\oper{\epsilon} \cdot \tilde{\vec{E}} + \frac{c^2}{\omega^2}\,
\vec{k} \times \Big[\oper{\mu}^{-1} \cdot (\vec{k} \times \tilde{\vec{E}}) \Big] = 0,
\end{gather}
in agreement with Maxwell's equations \cite[Sec.~3.4]{book:stix}. Similarly, varying $\mcc{L}$ with respect to $\tilde{\vec{E}}$ yields the complex-conjugate equation.

Alternatively, if the polarization vector $\vec{e}$ is prescribed (or considered as an independent field), one can as well introduce a scalar amplitude instead, say, $a = |\tilde{\vec{E}}|$. This yields $\mcc{L} = \mcc{D}(\omega, \vec{k})a^2$, with $\mcc{D}(\omega, \vec{k})$ given by
\begin{gather}\label{eq:dprime}
\mcc{D} = \frac{1}{16\pi} \left[\vec{e}^* \cdot \oper{\epsilon} \cdot \vec{e} - 
\frac{c^2}{\omega^2}\,(\vec{k} \times \vec{e}^*) \cdot \oper{\mu}^{-1} \cdot (\vec{k} \times \vec{e})
\right].
\end{gather}
The dispersion relation that follows [\Eq{eq:linL}] is \Eq{eq:drem} multiplied by $\vec{e}^*$.

%--------------------------------------
\subsection{Wave action and EMT}
\label{sec:kin}

\msection{Action} The action density $\mc{I}$ is now obtained straightforwardly by differentiating $\mcc{L}$ [\eg \Eq{eq:lagr1}] with respect to $\omega$:
\begin{gather}\notag
\mc{I} = \frac{1}{16\pi}\,\left[
\tilde{\vec{E}}^* \cdot \oper{\epsilon}_\omega \cdot \tilde{\vec{E}} 
+ \frac{2}{\omega}\,\tilde{\vec{H}}^* \cdot \tilde{\vec{B}} -
\tilde{\vec{B}}^* \cdot (\oper{\mu}^{-1})_\omega \cdot \tilde{\vec{B}}
\right],
\end{gather}
where we used $\tilde{\vec{H}}^* \cdot \tilde{\vec{B}} = \tilde{\vec{B}}^* \cdot \tilde{\vec{H}}$, due to $\oper{\mu}$ being Hermitian. From $\mcc{L} = 0$ [\Eq{eq:linL}], one also has
\begin{gather}\label{eq:equal}
\tilde{\vec{E}}^* \cdot \tilde{\vec{D}} = \tilde{\vec{H}}^* \cdot \tilde{\vec{B}}.
\end{gather}
Thus, $\mc{I} = \mc{I}^{(E)} + \mc{I}^{(B)}$, where
\begin{gather}
\mc{I}^{(E)} = \frac{1}{16\pi}\,\left[
\tilde{\vec{E}}^* \cdot \oper{\epsilon}_\omega \cdot \tilde{\vec{E}} + 
\frac{1}{\omega}\,\tilde{\vec{E}}^* \cdot \oper{\epsilon} \cdot \tilde{\vec{E}}\right], \\
\mc{I}^{(B)} = \frac{1}{16\pi}\,\left[
\frac{1}{\omega}\,\tilde{\vec{H}}^* \cdot \oper{\mu}\cdot \tilde{\vec{H}} -
\tilde{\vec{B}}^* \cdot (\oper{\mu}^{-1})_\omega \cdot \tilde{\vec{B}}
\right].
\end{gather}
One can further substitute
\begin{multline}
\tilde{\vec{B}}^* \cdot (\oper{\mu}^{-1})_\omega \cdot \tilde{\vec{B}} 
= \tilde{\vec{B}}^* \cdot (\oper{\mu}^{-1})_\omega \cdot \oper{\mu} \cdot \tilde{\vec{H}} = \\
= - \tilde{\vec{B}}^* \cdot \oper{\mu}^{-1} \cdot \oper{\mu}_\omega \cdot \tilde{\vec{H}}
= - \tilde{\vec{H}}^* \cdot \oper{\mu}_\omega \cdot \tilde{\vec{H}},
\end{multline}
where we used $(\oper{\mu}^{-1} \cdot \oper{\mu})_\omega \equiv 0$. Therefore, 
\begin{gather}\label{eq:emi0}
\mc{I} = \frac{1}{16\pi \omega}\,\left[
\tilde{\vec{E}}^* \cdot (\omega\oper{\epsilon})_\omega\cdot \tilde{\vec{E}} +
\tilde{\vec{H}}^* \cdot (\omega\oper{\mu})_\omega \cdot \tilde{\vec{H}} 
\right].
\end{gather}

\msection{Canonical EMT} The elements of the wave canonical EMT [\Eq{eq:block}] are readily obtained from \Eq{eq:emi0}. For completeness, we summarize them here once again:
\begin{gather}\label{eq:all}
\mc{E} = \omega \mc{I}, \quad \vec{\mc{Q}} = \vec{v}_{\rm g} \omega \mc{I}, \quad \vec{\mc{P}} = \vec{k}\mc{I}, \quad \oper{\Pi} = \vec{k} \vec{v}_{\rm g} \mc{I}.
\end{gather}

\msection{Kinetic EMT} Assuming that dissipation is negligible and the medium is isotropic, the wave kinetic EMT, as well as the kinetic angular momentum, can also be found, namely, using the results from \Sec{sec:amc}. In the general case, one can employ \Eqs{eq:dpdef} and \eq{eq:dedef}, substituting \Eq{eq:Phia} for $\Phi$. In the isotropic-fluid approximation, \Eqs{eq:vefin}, \eq{eq:pfin}, and \eq{eq:sigfin} can be used in combination with \Eqs{eq:emi0} and \eq{eq:all} taken in the medium rest frame. Due to \Eq{eq:mcUel}, one can also take, in particular,
\begin{gather}\label{eq:aux60}
\vec{\mcc{R}} = -\frac{\gamma^2 \vec{v}}{c^2} \left(
n\,\frac{\pd \epsilon}{\pd n}\,\frac{|\tilde{\vec{E}}|^2}{16\pi}
-n\,\frac{\pd\mu^{-1}}{\pd n}\,\frac{\tilde{|\vec{B}}|^2}{16\pi}
\right),
\end{gather}
where the expression in parenthesis (equal to the interaction-Lagrangian density $-\mc{U}$) is Lorentz-invariant. Hence $\vec{\mcc{R}}$ can be attributed to electrostriction and magnetostriction \cite{ref:dewar77, ref:robinson75}. Besides, one can show that Eqs.~(170) and (171) of \Ref{ref:dewar77}, derived there from different considerations, are recovered from our \Eqs{eq:vefin}, \eq{eq:pfin}, and \eq{eq:sigfin} as a special case. (The proof is straightforward and will not be presented here.) For practical ramifications of our results in application to electromagnetic waves also see \Ref{ref:dewar77} and \Sec{sec:disc}.

%--------------------------------------
\subsection{Dissipative waves}
\label{sec:emdiss}

In the presence of dissipation, the dispersion relation flowing from Maxwell's equations is similar to that in \Sec{sec:U}. Namely, it can be written as $\mcc{D}(\Omega, \vec{K}) = 0$, where $\mcc{D}$ has the same form as in \Eq{eq:dprime}, yet now with
\begin{gather}
\oper{\epsilon} = \oper{\epsilon}' + i\oper{\epsilon}'', 
\quad
\oper{\mu} = \oper{\mu}' + i\oper{\mu}''.
\end{gather}
where $\oper{\epsilon}'$ and $\oper{\mu}'$ are Hermitian, and $i\oper{\epsilon}''$ and $i\oper{\mu}''$ are anti-Hermitian. Using that
\begin{gather}
(\oper{\mu}' + i\oper{\mu}'')^{-1} \approx \oper{\mu}'^{-1} - i\oper{\mu}'^{-1}\cdot \oper{\mu}''\cdot \oper{\mu}'^{-1},
\end{gather}
we can hence write, for $\mcc{D}$ evaluated at real $(\omega,\vec{k})$, that $\mcc{D} = \mcc{D}' + i\mcc{D}''$, where $\mcc{D}'$ and $\mcc{D}''$ are real and given by
\begin{gather}
\mcc{D}' = \frac{1}{16\pi} \left[\vec{e}^* \cdot \oper{\epsilon}' \cdot \vec{e} - 
\frac{c^2}{\omega^2}\,(\vec{k} \times \vec{e}^*) \cdot \oper{\mu}'^{-1} \cdot (\vec{k} \times \vec{e})
\right],\notag
\end{gather}
\begin{multline}
\mcc{D}'' = \frac{1}{16\pi} \Bigg[\vec{e}^* \cdot \oper{\epsilon}'' \cdot \vec{e} \\ 
+ \frac{c^2}{\omega^2}\,(\vec{k} \times \vec{e}^*) \cdot \oper{\mu}'^{-1}\cdot \oper{\mu}''\cdot \oper{\mu}'^{-1} \cdot (\vec{k} \times \vec{e})
\Bigg].\notag
\end{multline}

According to \Sec{sec:dissw}, we can infer $\mc{I}$ directly from \Eq{eq:emi0} by replacing $\mcc{D}$ with $\mcc{D}'$, so
\begin{gather}
\mc{I} = \frac{1}{16\pi \omega}\,\left[
\tilde{\vec{E}}^* \cdot (\omega\oper{\epsilon}')_\omega\cdot \tilde{\vec{E}} +
\tilde{\vec{H}}^* \cdot (\omega\oper{\mu}')_\omega \cdot \tilde{\vec{H}} 
\right].
\end{gather}
Then the known formula \cite[Sec.~80]{book:landau8} for the energy density is recovered from $\mc{E} = \omega \mc{I}$. Other local properties of the wave are found from \Eqs{eq:eqppi2} and \eq{eq:meqdef}, the dissipation rate $\Gamma$ is found from \Eq{eq:Gamma}, and \Eq{eq:id1} yields
\begin{gather}\label{eq:wdiss}
\imath_{\rm loss} = \frac{1}{8\pi}\left(
\tilde{\vec{E}}^*\cdot\oper{\epsilon}''\cdot\tilde{\vec{E}}
+\tilde{\vec{H}}^*\cdot\oper{\mu}''\cdot\tilde{\vec{H}}
\right),
\end{gather}
where we substituted $\varkappa = 2$, since $A = a^2$. The expression for the dissipation power density, $w_{\rm loss} = \omega \imath_{\rm loss}$, hence also agrees with the known formula \cite[Sec.~80]{book:landau8}.

%--------------------------------------
\subsection{Dielectric media}
\label{sec:diel}

Since $\tilde{\vec{B}}$ is proportional to $\tilde{\vec{E}}$, one usually can define the high-frequency medium-response tensors $\oper{\epsilon}$ and $\oper{\mu}$ such that $\oper{\mu} = 1$ (in a selected frame of reference). As this is done often, \eg in plasma physics \cite{book:stix}, let us also simplify some of the above expressions for this particular case. First of all, \Eq{eq:lem} yields
\begin{gather}
\mcc{L} = \frac{1}{16\pi} \left[\tilde{\vec{E}}^* \cdot \oper{\epsilon}'\cdot \tilde{\vec{E}} - \frac{c^2}{\omega^2}\,|\vec{k} \times \tilde{\vec{E}}|^2\right],
\end{gather}
or $\mcc{L} = \mcc{L}^{(0)} +\tilde{\vec{E}}^* \cdot \oper{\chi}' \cdot \tilde{\vec{E}}/(16\pi)$, where $\mcc{L}^{(0)}$ is the vacuum Lagrangian [\Eq{eq:l02}], and we introduced the electric susceptibility $\oper{\chi} \doteq \oper{\epsilon} - 1$. Then the wave energy~is
\begin{gather}
\mc{E} = \frac{1}{16\pi}\,\left[\tilde{\vec{E}}^* \cdot (\omega \oper{\epsilon}')_\omega \cdot \tilde{\vec{E}} +|\tilde{\vec{B}}|^2\right],
\end{gather}
or, equivalently [due to $\tilde{\vec{E}}^* \cdot \oper{\epsilon}' \cdot \tilde{\vec{E}} = |\tilde{\vec{B}}|^2$; cf. \Eq{eq:equal}],
\begin{gather}\label{eq:energy22}
\mc{E} = \frac{1}{16\pi\omega}\,\tilde{\vec{E}}^* \cdot (\omega^2\oper{\epsilon}')_\omega \cdot \tilde{\vec{E}}.
\end{gather}
Also, as usual, the canonical momentum density equals
\begin{gather}\label{eq:PPP}
\vec{\mc{P}} = \vec{k}\mc{E}/\omega.
\end{gather}
One can show, using \Eq{eq:drem}, that the latter is just a more concise form of the corresponding expression in \Ref{tex:bers75}. Contrary to \Ref{ref:cary80}, calculated there is thus not the total, but only the canonical momentum (and the canonical energy) of the wave; see also \Refs{ref:best65, ref:kholmetskii11}.

Following \Ref{book:stix}, let us also separate the energy flux $\vec{\mc{Q}}$ into the electromagnetic part and the kinetic part. Specifically, using $\vec{\mc{Q}} = - \omega\mcc{L}_\vec{k}$, one can write it as $\vec{\mc{Q}} = \vec{\mc{S}} + \vec{\mc{K}}$, where $\vec{\mc{S}} = - \omega\mcc{L}^{(0)}_{\vec{k}}$, and 
\begin{gather}
\vec{\mc{K}} = - \frac{\omega}{16\pi}\,\tilde{\vec{E}}^* \cdot \oper{\chi}_\vec{k}' \cdot \tilde{\vec{E}}.
\end{gather}
The latter is recognized as the energy flux density caused by the presence of the medium \cite[Chap.~4]{book:stix}, whereas
\begin{align}
\vec{\mc{S}} 
& = \frac{c^2}{16\pi\omega} \Big\{
(\underline{\vec{k}} \times \tilde{\vec{E}}^*) \cdot (\vec{k} \times \tilde{\vec{E}}) +
(\vec{k} \times \tilde{\vec{E}}^*) \cdot (\underline{\vec{k}} \times \tilde{\vec{E}})
\Big\}_\vec{k} \notag\\
& = \frac{c^2}{16\pi\omega} \Big\{
\underline{\vec{k}} \cdot \big[\tilde{\vec{E}}^* \times (\vec{k} \times \tilde{\vec{E}})\big]
+
\underline{\vec{k}} \cdot \big[\tilde{\vec{E}} \times (\vec{k} \times \tilde{\vec{E}}^*)\big]
\Big\}_\vec{k} \notag\\
& = \frac{c^2}{16\pi\omega} \Big\{
\tilde{\vec{E}}^* \times (\vec{k} \times \tilde{\vec{E}})
+
\tilde{\vec{E}} \times (\vec{k} \times \tilde{\vec{E}}^*)
\Big\} \notag\\
& = \frac{c}{8\pi} \,\text{Re}\,(\tilde{\vec{E}} \times \tilde{\vec{B}}^*)
\end{align}
is the time-averaged Poynting vector, \ie the ``vacuum part'' of $\vec{\mc{Q}}$. [Here we substituted \Eq{eq:l02} and used underlining to specify where the differentiation applies.] Recalling that $\vec{\mc{Q}} = \mc{E}\vec{v}_{\rm g}$, one then also recovers the known formula \cite[Chap.~4]{book:stix}
\begin{gather}\label{eq:vgsk}
\vec{v}_{\rm g} = (\vec{\mc{S}} + \vec{\mc{K}})/\mc{E}.
\end{gather}
Below, several examples of specific dielectrics will be discussed to illustrate these and earlier formulas.

%--------------------------------------
\subsection{Examples}
\label{sec:examples}

\msection{Waves in fluids at rest} Let us first summarize, using the results of \Sec{sec:diel}, the densities of the kinetic energy, momentum, and angular momentum for a wave in a fluid dielectric at rest:
\begin{gather}\notag
\varepsilon = \mc{E}, \quad \vec{\rho} = (\vec{\mc{S}} + \vec{\mc{K}})/c^2,  \quad \vec{\mu} = [\vec{x} \times (\vec{\mc{S}} + \vec{\mc{K}})]/c^2.
\end{gather}
Electromagnetic waves in vacuum can be considered as a special case and have $\omega^2 = c^2k^2$, so $\vec{v}_{\rm g} = c^2\vec{k}/\omega = c \Bbbk$, where ${\Bbbk \doteq \vec{k}/k}$. Then $\mc{E} = |\tilde{E}|^2/(8\pi)$; $c^2\vec{\mc{P}}$ and $\vec{\mc{Q}}$ are equal to each other (so the EMT is symmetric, and canonical quantities coincide with kinetic quantities) and $\vec{\mc{S}} = \Bbbk c\mc{E}$; also, $\oper{\Pi} = \Bbbk\Bbbk\,\mc{E}$ equals minus the time-averaged Maxwell stress tensor. Thus, in this case, the wave EMT coincides with the electromagnetic stress-energy tensor \cite[Sec.~32]{book:landau2}. Besides, $\vec{\mc{M}}$ and $\vec{\mu}$ are both equal to ${\vec{x} \times \vec{\mc{S}}/c^2}$, in agreement with the traditional definition of the wave angular momentum density in vacuum \cite{ref:allen92}.

\msection{Electrostatic waves in beams and plasmas} To also illustrate waves in moving dielectrics, consider further a relativistic electron beam with electrostatic oscillations seeded on it with~$\vec{k}$ parallel to the beam velocity~$\vec{v}$. Assuming that the beam is cold, it acts as a fluid medium isotropic in its rest frame. (In fact, as long as the dynamics is one-dimensional, having isotropy is inessential.) Since there is no dissipation in a cold beam, we revert here to using primes as a reference to that frame. The proper ponderomotive potential, same for all electrons, can then be written as $\Phi' = e^2|\tilde{E}|^2/(4m_{\rm e} \omega'^2)$ \cite{my:acti}, where $e$ and $m_{\rm e}$ are the electron charge and mass; remember also that the longitudinal field satisfies $\tilde{E} = \tilde{E}'$. Then $\epsilon' = 1 - \omega_{\rm p}'^2/\omega'^2$ (cf. \Sec{sec:U}), where $\omega_{\rm p}' \doteq (4\pi n'_{\rm e} e^2/m_{\rm e})^{1/2}$, and $n'_{\rm e}$ is the beam proper density. This yields the dispersion relation in the form $\omega'^2 = \omega_{\rm p}'^2$, and $v_{\rm g}' = 0$ in particular. From \Eq{eq:energy22} we hence get $\mc{E}' = |\tilde{E}|^2/(8\pi)$, so $\mc{U}' = \mc{E}'/2$ [cf. \Eq{eq:mcUdef}]; then \Eqs{eq:vefin} and \eq{eq:pfin} lead~to
\begin{gather}\label{eq:epbeam}
\varepsilon = (3\gamma^2 - 1)\,\frac{|\tilde{E}|^2}{16\pi}, \quad \rho = \frac{3\gamma^2 v}{c^2}\,\frac{|\tilde{E}|^2}{16\pi}.
\end{gather} 
Since $\omega' = \gamma (\omega - kv)$, the dispersion relation in the laboratory frame is $\omega = \omega_{\rm p}'/\gamma + kv$, and $v_{\rm g} = v$; therefore, $\rho = [1 - 1/(3\gamma^2)]^{-1} \varepsilon v_{\rm g}/c^2$, or $\rho \approx 3 \varepsilon v_{\rm g}/(2c^2)$ at $\gamma \approx 1$. This differs by a factor $3/2$ from Abraham's result (expected at nonrelativistic velocities), but the discrepancy can be readily explained. Recall that Abraham's limit [\Eq{eq:abrappr}] is derived by neglecting $v$ but retaining $v_{\rm g}$. For the electrostatic waves in question, this is legitimate only when $v$ equals zero exactly; in that case, just like Abraham's formula, \Eq{eq:epbeam} yields $\rho = 0$. Note that this result is also understood from the fact that, at $v = 0$, each nonrelativistic electron has a zero average momentum, while the instantaneous momentum of the electrostatic \textit{field} (proportional to the instantaneous Poynting vector \cite[Sec.~6.9]{book:jackson}) is zero identically.

Now consider thermal effects, which will render beam oscillations identical to Langmuir waves in warm collisionless plasma \cite[Chap.~8]{book:stix}. We can take the electron ponderomotive potential in the form $\Phi = e^2|\tilde{E}|^2/[4m_{\rm e} (\omega - ku)^2]$, where $u$ is the average velocity of an individual particle; then the general Bohm-Gross dispersion relation is recovered \cite{my:acti}. As usual, we assume that the electron thermal velocity $v_{\rm T}$ satisfies $\xi_0 \doteq \omega/(kv_{\rm T}) \gg 1$, and the electron average velocity is zero, so one gets  $\omega^2 \approx \omega_{\rm p}^2 + 3k^2v_{\rm T}^2$ and $v_{\rm g} \approx 3kv_{\rm T}^2/\omega$. (See \Ref{ref:bohm49} for how to deal with the singularity at $u = \omega/k$.) The canonical energy and momentum densities are then found to be $\mc{E} \approx (\omega_{\rm p}/\omega)^2|\tilde{E}|^2/(8\pi)$ and $\mc{P} = k\mc{E}/\omega$. This time we yet cannot apply Abraham's formula [\Eqs{eq:abrappr}] to further find $\varepsilon$ and $\rho$, because, as an ensemble of electrons with different velocities, warm plasma is not a simple fluid. Thus we revert to \Eqs{eq:dpdef} and \eq{eq:dedef}, which give
\begin{gather}
\Delta\varepsilon \approx - \frac{3 k^2 v_{\rm T}^2}{\omega_{\rm p}^2}\,\frac{|\tilde{E}|^2}{8\pi},
\quad
\Delta \rho \approx -\frac{k}{\omega}\,\frac{\omega_{\rm p}^2}{\omega^2}\,\frac{|\tilde{E}|^2}{8\pi}.
\end{gather} 
Hence, \Eqs{eq:fullpm} finally yield $\varepsilon = |\tilde{E}|^2/(8\pi)$ and $\rho = 0$, at least up to terms of the order of $\xi_0^{-4}$. 

Since $\rho$ is independent of the amplitude, the evolution of Langmuir waves does not affect the electron average momentum (zero in our example). This agrees with the standard quasilinear theory \cite{book:krall}; however, an explanation is due regarding the details. Notice that a Langmuir envelope seems to transport canonical momentum due to nonzero $v_{\rm g}$, but does not seem to transport any kinetic momentum, as $\rho$ is zero; on the other hand, one may expect that the total momentum in any volume enclosing the pulse is defined unambiguously in quiescent plasma, so one arrives to paradox, much like the AMC. To resolve this, recall that above we assumed zero average velocity \textit{inside} the wave, so the outside plasma had to be flowing, as the amplitude gradient at the interface causes ponderomotive acceleration. Accordingly, to keep the outside velocity zero, we must allow for a nonvanishing flow of electrons inside the pulse, and that is precisely where the total momentum is stored. In other words, Langmuir waves do transport momentum through ambient plasma, just like other waves. This, if anything, may serve as a particularly vivid illustration for how the abstract resolution of the AMC relates to experiment.

\msection{Kinetic waves} The small but finite $\Delta \varepsilon/\mc{E} \sim \xi_0^{-2}$ that we found for Langmuir waves is a purely kinetic effect, which does not fit into the resting-fluid model, \Eq{eq:primedep}. For other waves in collisionless plasma, having finite $\xi_0$ generally renders the fluid model inapplicable too \cite{book:stix}. (Same applies to collisionless gas \cite{ref:stubbe94}, but collisional media are more forgiving.) Light waves in particular are affected at relativistic temperatures, which are possible in astrophysical settings and also in laboratory, say, in counterpropagating relativistic beams. But even more easily the fluid approximation can be broken for waves tuned in resonance with medium natural oscillations. Replacing $\xi_0$, the controlling parameters in this case are each of $\xi_\ell \doteq (\omega - \ell \Omega)/(k v_{\rm T})$, where $\Omega$ is the natural frequency, and $\ell$ is an arbitrary integer. This is well-known for a cyclotron resonance \cite{book:stix}, but interactions at other, even quantum resonances are similar \cite{my:nlinphi}. For example, one can show that $\Delta\varepsilon/\mc{E} \sim \xi_1^{-1}$ when a medium, resting on average, consists of two counterpropagating low-density beams of nonrelativistic oscillators with $\omega \approx \Omega$. However, elaborating on this is beyond the scope of our paper, and our intention here is only to re-emphasize that Abraham's formulas for $\varepsilon$ and $\vec{\rho}$ may not apply even when the medium average velocity is zero.

%--------------------------------------
\subsection{Quantum interpretation}
\label{sec:quantum}

Let us now recast our findings in the photon language. To do so, however, we first need to recall what a photon actually is, and we start with one in vacuum. 

A vacuum photon is defined as an elementary excitation of electromagnetic field, in a certain energy eigenstate that determines both the spatial structure of the mode and also its frequency. For example, the mode can be a standing wave with a certain number of nodes in a finite-size box. Yet if the box size is large enough or infinite (as it would be for free field), the energy gap between neighboring modes is negligible, so we can loosen the definition. Specifically, neighboring modes can then be excited coherently such that their interference produces a propagating envelope with a size negligible for our purposes. Precisely these envelopes we call photons, much like it is often done for regular particles in quasiclassical theories \cite{book:ashcroft}. Hence the GO approximation is also adopted automatically.

Now that we have introduced a photon in vacuum, let us define one in a dispersive medium. It is natural to continue thinking of a photon as a quasiparticle in this case, \ie an object that is \textit{conserved} within the GO approximation. To understand what it means, consider a vacuum photon entering a dispersive medium. We will assume no scattering at the boundary, to ensure that GO remains valid; \ie (i)~there is no reflection, and (ii)~the wave continues inside the medium as a quasimonochromatic field. In other words, only one of the internal eigenwaves, or branches of the dispersion relation, is excited, while others are nonresonant to the incident photon and thus remain quiescent. In the general case, when the medium is  both inhomogeneous and time-dependent, there is exactly one GO integral: the wave action. Therefore, it is the action conservation that must be associated with the photon conservation. And since all the action belongs to a single branch, so must all photons.

As one moves away from the boundary, the initial branch can adiabatically transform into something very different from the original electromagnetic wave. In that case it may be more natural to assign a different name to its elementary excitations, \ie call them not photons but, say, plasmons or polaritons. Remember, however, that a wave with given $\omega$ and $\vec{k}$ can be associated with only one type of quasiparticles, whatever they are called, and those must account for oscillations of \textit{both} the field and the medium. (In other words, GO photons cannot be separated from GO polaritons in principle, as long as a single branch of the dispersion curve is considered.) Absent a better term \cite{foot:dressed}, we thus adopt ``photon'' as the generic term for all such elementary excitations. 

Much like an electron interacting with a vector potential, a photon defined this way can hence be assigned two different momenta. The first, canonical momentum $\vec{P}$ is the one that enters Hamilton's equations [\Eqs{eq:phm2}] and is due to the corresponding Noether symmetry \textit{of the wave subsystem}. It is given by $\vec{P} = \hbar \vec{k}$; surprisingly, this expression holds in any dissipationless medium whatsoever and even for nonlinear waves. (Not so for the canonical energy $\hbar\omega$.) The second, kinetic momentum $\vec{p}$ is defined as the amplitude-dependent part of the average momentum of the whole physical system divided by the total number of photons. It is due to the WMS Lorentz invariance, or, in other words, the corresponding Noether symmetry \textit{of the whole system}, which includes the medium. In particular, for isotropic fluid medium at rest we show that $\vec{p} = \hbar\omega\vec{v}_{\rm g}/c^2$, in agreement with Abraham \cite{foot:history}. Remember, however, that this result is less general than Minkowski's formula for $\vec{P}$; for example, it does not apply in moving media (such as beams) and may not hold at relativistic temperatures.

Note, finally, that the per-photon average momenta of fields and particles taken separately are not associated with any conservation laws, in contrast with $\vec{P}$ and $\vec{p}$. Therefore they cannot be attributed to any conserved quasiparticles and, in this sense, are less meaningful.

%%%%%%%%%%%%%%%%%%%%%%%%%%%%%%%%%%%%%%%
\section{Discussion}
\label{sec:disc}

In this paper, we pose classical GO axiomatically within the field-theoretical approach, while extending it to account for dissipation. The concept of a photon in a dielectric medium is introduced, and photon properties are calculated unambiguously. In particular, the canonical and kinetic momenta and angular momenta carried by a photon, as well as the two corresponding EMTs, are derived from first principles of Lagrangian mechanics. Responding to the questions posed in \Ref{ref:leonhardt06b}, we thus resolve the Abraham-Minkowski controversy pertaining to the definition of the photon energy-momentum and spin, clarify the applicability of Minkowski's and Abraham's formulas, and find corrections to them for various media (including cold, warm, and relativistic media).

Furthermore, the axiomatic formalism that we adopt leads also to other new results, since it applies not just to electromagnetic waves but to any linear waves in any dispersive media, including ones yet to be discovered. For example, the EMTs of acoustic waves follow, plus the pho\textit{n}on spin --- and these are only some of the AMC-related issues that otherwise remain under debate \cite{foot:mcdonald, ref:minasyan11, ref:levine62}. We show, in fact, that all wave mechanical properties flow from little more than the wave definition, whereas the specific internal physics of a medium is largely irrelevant. Since definitions are not really a matter of verification, experimental resolution of the AMC is obviated; at least, one may call into question exactly what aspect of the AMC an experiment might resolve. In addition, derivations of the EMTs and angular momenta for specific waves are obviated too, as they are subsumed under our more general theory. For instance, (i)~the canonical EMT that we present is general (\Sec{sec:linear}), (ii)~the force produced by a wave on a fluid-like medium is independent of the precise constitution of the medium (\Sec{eq:kinemtfl}), and (iii)~even understanding of kinetic waves does not require solving any field equations (\Sec{sec:examples}).

Let us emphasize, finally, that these findings have been made possible by our adopting the abstract Lagrangian formulation, whose utility for understanding general linear waves should hence be obvious. We also suggest \Refs{my:langact, my:mquanta, my:dense, arX:mycoin} as recent illustrations of how advantageous it is for analyzing the linear wave dynamics in plasmas. In addition, the \textit{nonlinear} Lagrangian theory has been getting a new spin recently, namely, in the context of plasma waves carrying autoresonantly trapped particles \cite{my:bgk, my:acti, my:actii, my:actiii, ref:khain12}. Those waves are unique in the sense that the trapped-particle nonlinearity is, within a certain range of parameters, independent of the wave amplitude or even gets strengthened when the amplitude decreases. Hence the traditional intuition and standard perturbative approaches often fail when applied to such waves, whereas the axiomatic GO not only holds but also offers the advantage of tractability \cite{my:actiii}. In particular, understanding how the wave momentum relates to the plasma Lagrangian yields precise quantitative predictions of (quite nontrivial) evolution of waves with trapped particles in nonstationary plasma \cite{tex:mytrcomp}. Thus, with this paper, we would like to attract attention to the axiomatic GO itself as a remarkably convenient framework for analyzing the wave basic physics, not only in the AMC context but also in the context of solving practical problems.

The work was supported by the NNSA SSAA Program through DOE Research Grant No. DE274-FG52-08NA28553 and by the U.S. DOE through Contract No. DE-AC02-09CH11466.

\gap
%%%%%%%%%%%%%%%%%%%%%%%%%%%%%%%%%%%%%%%
\appendix

\section{Auxiliary function ${\vec{G}^\lambda}_\mu$}

Here, we summarize the properties of a dimensionless matrix function ${\vec{G}^\lambda}_\mu$ introduced in \Eq{eq:gbardef}. First of all, notice an obvious equality
\begin{gather}\label{eq:linv}
{(\Lambda^{-1})^\lambda}_\nu(\vec{v}) = {\Lambda^\lambda}_\nu(-\vec{v}),
\end{gather}
which can also be checked by confirming that
\begin{gather}
{\Lambda^\mu}_\lambda(-\vec{v}){\Lambda^\lambda}_\nu(\vec{v}) = \delta^\mu_\nu.
\end{gather}
Then a direct calculation yields
\begin{gather}
{G^0}_{0l} = 0, \\
{G^i}_{0l} = {\Lambda^i}_l,\quad
{G^0}_{il} = \eta_{ij}{\Lambda^j}_l,\\
{G^i}_{jl} = (\delta^i_l v_j - \eta_{jl}v^i)(\gamma/c)/(\gamma+ 1),
\end{gather}
where we introduced the notation
\begin{gather}
{G^\lambda}_{\mu l} \equiv ({\vec{G}^\lambda}_\mu)_l \doteq 
(c/\gamma){(\Lambda^{-1})^\lambda}_\nu\, (\pd{\Lambda^\nu}_\mu/\pd v^l).
\end{gather}
(In particular, notice that the three $l$-components, ${G^\nu}_{\mu l}$, at $\vec{v} = 0$ happen to be the well-known Lorentz boost generators.) Let us now \textit{define} the function
\begin{gather}\label{eq:gdeflow}
G_{\nu\mu l}\doteq g_{\nu\lambda}{G^\lambda}_{\mu l}.
\end{gather}
Due to \Eqs{eq:metricgeta}, one finds the latter to be
\begin{gather}
G_{00l} = 0, \label{eq:G1}\\
G_{0il} = - G_{i0l} = - \eta_{ij}{\Lambda^j}_l,\label{eq:G23}\\
G_{ijl} = (\eta_{il}v_j - \eta_{jl}v_i)(\gamma/c)/(\gamma + 1),\label{eq:G4}
\end{gather}
so, in particular,
\begin{gather}
\vec{G}_{\nu\mu} = - \vec{G}_{\mu\nu}, \quad
(\vec{G}_{\mu\nu})_l \equiv G_{\mu\nu l}.
\end{gather}
Hence \Eq{eq:aux57} becomes
\begin{multline}
\pd_\vec{v} \mcc{L}' = 
- \gamma{\vec{G}^\lambda}_\mu g_{\lambda\nu}\mc{T}'^{\nu\mu}/c = - \gamma g_{\nu\lambda}{\vec{G}^\lambda}_\mu \mc{T}'^{\nu\mu}/c \\
= - \gamma\vec{G}_{\nu\mu} \mc{T}'^{\nu\mu}/c = \gamma \vec{G}_{\mu\nu}\mc{T}'^{\nu\mu}/c,
\end{multline}
which is exactly \Eq{eq:aux58}, where we substituted \Eq{eq:G1} and introduced
\begin{gather}\notag
\mcc{P}_l \doteq (\gamma/c)(G_{i0l}\mc{T}'^{0i} + G_{0il}\mc{T}'^{i0}), \quad \mcc{B}_l \doteq (\gamma/c)G_{ijl}\mc{T}'^{ji}.
\end{gather}
Finally, due to \Eqs{eq:G23} and \eq{eq:G4},
\begin{multline}
\mcc{P}_l = \gamma(\eta_{ij}{\Lambda^j}_l \mc{E}' v_{\rm g}'^i/c^2 - \eta_{ij}{\Lambda^j}_l \mc{P}'^i) \\
= \gamma{\Lambda^j}_l (\mc{E}' v'_{gj}/c^2 - \mc{P}'_j)
= [\gamma\oper{\Lambda} \cdot (\mc{E}' \vec{v}'_{\rm g}/c^2 - \vec{\mc{P}}')]_l,\notag
\end{multline}
\vspace{-10pt}
\begin{multline}
[(\gamma + 1)c^2/\gamma^2]\mcc{B}_l\\ 
= (\eta_{il}v_j - \eta_{jl}v_i)\mc{P}'^j v_{\rm g}'^i 
= (v'_{gl} v_j \mc{P}'^j - \mc{P}'_lv_i v_{\rm g}'^i) \notag\\
= [\vec{v}'_{\rm g} (\vec{v} \cdot \vec{\mc{P}}') - \vec{\mc{P}}' (\vec{v} \cdot \vec{v}'_{\rm g})]_l
= [\vec{v} \times (\vec{v}'_{\rm g} \times \vec{\mc{P}}')]_l,
\end{multline}
whence \Eqs{eq:mccP} and \eq{eq:mccB} readily follow.

%\bibliography{main,foot}

\begin{thebibliography}{100}
\bibitem{foot:veca}
Regardless of how $a$ is defined in a specific case [including whether $a$ is a
  scalar or a vector, in which case the polarization is also inferred
  consistently from \Eq{eq:L}], the resulting equations will be equivalent.
  Also, for linear waves, it is often convenient instead to have as independent
  variables the two complex-conjugate envelopes, and this does not affect our
  analysis either.
\bibitem{foot:xik}
The independent functions are $(a, \theta)$ but not $(a, k_\mu)$ because the
  latter, albeit describing the envelope, do not completely determine the wave
  field.
\bibitem{foot:manley}
The ACT also permits generalization to resonant interactions. For example, the
  beat phase $\Delta \theta \doteq \theta_1 - \theta_2$ of two resonant waves
  (denoted as 1 and 2) is a slow variable, so it may enter the GO Lagrangian
  density; \ie $\mcc{L} = \mcc{L}(a_1, a_2, k_{1\mu}, k_{2\lambda}, \Delta
  \theta; x^\nu)$. Then, $\delta_{\theta_1} S = 0$ yields
  ${\mc{J}_1^\mu}_{;\mu} = \mcc{L}_{\Delta \theta}$, and $\delta_{\theta_2} S =
  0$ yields ${\mc{J}_2^\mu}_{;\mu} = - \mcc{L}_{\Delta \theta}$. Hence the
  total action is conserved, $(\mc{J}_1^\mu + \mc{J}_2^\mu)_{;\mu} = 0$, which
  is a reincarnation of the Manley-Rowe theorem~\cite{my:manley,
  ref:brizard95}.
\bibitem{foot:sign}
Notice that we define the EMT sign in a non-conventional way so the expressions
  be simplified for the adopted metric signature.
\bibitem{foot:noether}
For $g_{ij} = \text{diag}\,(-1,1,1,1)$ and a Lagrangian density $\mc{L}(\psi,
  \psi_{,\mu})$, the EMT \cite{foot:sign}, satisfying
  ${{T_\alpha}^\beta}_{,\beta} = 0$, is ${T_\alpha}^\beta \doteq -
  \psi_{,\alpha}\,\mc{L}_{\psi_{,\beta}} + \delta^\beta_\alpha \mc{L}$
  \cite[Sec.~32]{book:landau2}. In particular, this explains why \Eq{eq:emt2}
  is expected from \Eq{eq:L}, since ${\mcc{L}_{a_{,\beta}} \equiv 0}$.
\bibitem{foot:mcin}
Mind that $\mc{T}^{\alpha\beta}$ is the EMT of the wave rather than of the
  whole system, so it is often perceived solely as a mathematical abstraction
  and denied having a physical meaning \cite{ref:mcintyre81}. However, the very
  existence of such a thing as the ``elementary physical meaning''
  \cite{ref:mcintyre81} is questionable in the general case, and, in a sense,
  the properties of the wave ``physical'' EMT render it even less real than the
  however abstract $\mc{T}^{\alpha\beta}$ (cf. \Sec{sec:amc}). See also
  \Ref{book:hehl}.
\bibitem{foot:meff}
Then the particle canonical mass [\Eq{eq:meff}] is expected to be
  medium-dependent, as indeed found in solid state theory \cite{book:ashcroft}
  and also for ponderomotively-driven classical particles \cite{my:mneg,
  my:trieste08, my:meffham}.
\bibitem{foot:singo}
The singularity of a circularly polarized field is also understood from the
  fact that its polarization ellipse has degenerate orientation. For details
  see \Ref{ref:freund02} and papers on singular optics cited therein.
\bibitem{foot:elem}
Particles are assumed elementary in the sense that their momenta are determined
  entirely by the translational velocities. This assumption is invalid for
  calculating relativistic corrections to $\Delta\vec{\rho}$ when particles are
  anisotropic. Indeed, an anisotropic medium experiences a torque from the wave
  when $\vec{v}_{\rm g}$ is not parallel to $\vec{k}$ (\Sec{sec:angmom}). As
  the angular momenta of particles change then, their relativistic centers of
  mass \cite{ref:dixit82, book:moller} drift and thus carry an additional
  momentum not included in our model.
\bibitem{foot:dressed}
Sometimes the term ``dressed photon'' is used instead; see, \eg
  \Ref{ref:cascio92}. We, however, prefer not to use it here because dressed
  are the \textit{only} photons that can be defined classically within GO. An
  exception is a medium which has a weak effect on the wave dispersion; for
  some details see \Refs{my:nlinphi, my:kchi}.
\end{thebibliography}

\end{document}